\documentclass[twocolumn,prb]{revtex4}

\usepackage{graphicx}
\usepackage{epsfig}
\usepackage{amsmath, amssymb}
\usepackage{verbatim}
\usepackage{bm}

\def\v{\vec}
\renewcommand{\b}[1]{\boldsymbol{#1}}
\newcommand{\rchi}{{\mathpalette\irchi\relax}}
\newcommand{\irchi}[2]{\raisebox{\depth}{$#1\chi$}}
\newcommand{\rgamma}{{\mathpalette\irgamma\relax}}
\newcommand{\irgamma}[2]{\raisebox{\depth}{$#1\gamma$}}
\newcommand{\gammabar}{\ensuremath\gamma\kern-0.53em-}

\begin{document}

\title{Theory of defects in Abelian topological states}
\author{Maissam Barkeshli, Chao-Ming Jian, and Xiao-Liang Qi}
\affiliation{Department of Physics, Stanford University, Stanford, CA 94305 }

\begin{abstract}
The structure of extrinsic defects in topologically ordered states of matter is host to a rich set of universal physics.
Extrinsic defects in 2+1 dimensional topological states include line-like defects, such as boundaries between
topologically distinct states, and point-like defects, such as junctions between different line defects.
Gapped boundaries in particular can themselves be \it topologically \rm distinct, and the junctions between
them can localize topologically protected zero modes, giving rise to topological
ground state degeneracies and projective non-Abelian statistics. In this paper, we develop a general
theory of point defects and gapped line defects in 2+1
dimensional Abelian topological states. We derive a classification of topologically
distinct gapped boundaries in terms of certain maximal subgroups of quasiparticles with mutually
bosonic statistics, called Lagrangian subgroups. The junctions between different
gapped boundaries provide a general classification of point defects in topological states,
including as a special case the twist defects considered in previous works.
We derive a general formula for the quantum dimension of these point defects, a general
understanding of their localized ``parafermion'' zero modes, and we define a notion of
projective non-Abelian statistics for them. The critical phenomena between topologically distinct gapped boundaries can be understood in
terms of a general class of quantum spin chains or, equivalently, ``generalized parafermion''
chains. This provides a way of realizing exotic 1+1D generalized parafermion conformal field
theories in condensed matter systems.
\end{abstract}

\maketitle

\tableofcontents

\section{Introduction}

One of the most fundamental discoveries in condensed matter physics has been the
understanding of topologically ordered states of matter.\cite{wen04,nayak2008} Topologically ordered states
are gapped many-body states that possess quasiparticle excitations with fractional statistics and fractional charges,
topology-dependent ground state degeneracies, different patterns of
long-range entanglement\cite{levin2006,kitaev2006b}, and many other exotic characteristics. The most common
topological states seen experimentally are the fractional quantum Hall (FQH) states.
There is also increasing support from numerical simulations and experiments\cite{balents2010,lee2008,jiang2008,depenbrock2012,yan2011,jiang2012a,wang2011,jiang2012b}
that topologically ordered states are found in frustrated magnets.

\begin{figure}
\centerline{
\includegraphics[width=2.3in]{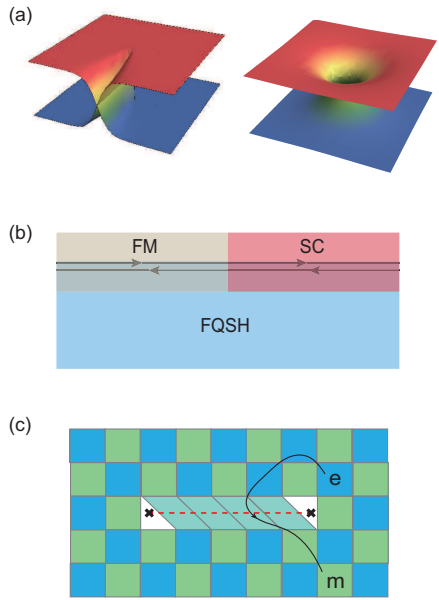}
}
\caption{Examples of point defects studied previously in the literature.
(a) Genons in bilayer systems\cite{barkeshli2012a,barkeshli2013genon} (b) domain walls between ferromagnetic
and superconducting backscattering at the edge of a fractional quantum spin Hall (FQSH)
state,\cite{clarke2013,lindner2012,cheng2012} (c) lattice dislocations in solvable models of $Z_N$ topological order.\cite{bombin2010,you2012}
\label{fig:genon}
}
\end{figure}

Recently, a new direction in the study of topologically ordered states, called twist
defects, or extrinsic defects, has attracted increasing research interest.\cite{barkeshli2010,bombin2010,
barkeshli2012a,barkeshli2013genon,barkeshli2013,you2012,you2013,clarke2013,lindner2012,cheng2012,vaezi2013,
brown2013,kitaev2012,mesaros2013,fuchs2012,levin2013,wang2012,beigi2011,kapustin2010,kapustin2011}
An extrinsic defect is a point-like or line-like defect either in a topological state, or on the interface between two topologically distinct
states, which leads to new topological properties that are absent in the topological state without the defect.
A crucial distinction between extrinsic defects and the more familiar quasiparticle excitations is that the former are not
deconfined excitations of the system; rather, the energy cost for separating point defects, for instance,
will generally depend either logarithmically or linearly on their separation.

A simple example of an extrinsic defect is the ``genon" defined in Ref. \onlinecite{barkeshli2012a,barkeshli2013genon}.
As shown in Fig. \ref{fig:genon}, a branch-cut line is introduced in a bilayer topological state, across which the two
layers are exchanged.\cite{barkeshli2010} The genon in this case is defined as the branch-cut point where the branch-cut line ends. From its
definition one can see that a bilayer system with genons is topologically equivalent to a single layer system on
a Riemann surface.\cite{barkeshli2010,barkeshli2012a} Using such a mapping,
the topological properties of genons such as quantum dimension and
(projective) braiding statistics can be studied systematically.\cite{barkeshli2013genon} Even when the topological state in each layer is an
Abelian theory, the genons have non-Abelian statistics. This has led to a recent experimental proposal for
synthesizing a wide variety of possible topological qubits using the simplest Abelian bilayer FQH states.\cite{barkeshli2013}
It has also been shown that the braiding statistics of genons can allow for universal topological quantum computation (TQC) even in cases
where the host topological state without the genons is not by itself universal for TQC.\cite{barkeshli2013genon}

Extrinsic defects with the same type of non-Abelian statistics as some of the genons studied in Ref. \onlinecite{barkeshli2012a,barkeshli2013}
have also been proposed in other physical systems. These include lattice defects in certain exactly solvable $Z_N$ rotor models,\cite{you2012,you2013}
FQH states in proximity with superconductivity (SC), and fractional quantum spin Hall (FQSH) states in proximity with SC and ferromagnetism
 (FM). \cite{lindner2012,clarke2013,cheng2012} The latter FQSH proposals are generalizations of earlier proposals
of realizing Majorana zero modes on the boundary of the quantum spin Hall insulator\cite{fu2009d}. In these FQSH
realizations, the extrinsic defect is a point on the boundary of the system where the boundary condition changes,
while the rest of the boundary is in a gapped state.  For Abelian states, the extrinsic defects realized in the
superconducting proximity proposals reviewed above can be mapped to genons in suitable bilayer
Abelian states\cite{barkeshli2013genon}.

In this paper, we develop a general theory of extrinsic defects in Abelian topological states, which generalize the
extrinsic defects reviewed above to the most generic possible form. For two-dimensional topological states,
there are two general forms of extrinsic defects. These are line defects, which separate two different
or identical topological states, and point defects, which may live in a topological state (such as genons)
or live on line defects (such as the FM/SC domain wall on the FQSH edge). We demonstrate that all extrinsic defects can
be mapped to {\it boundary defects}, {\it i.e.}, boundary lines of topological states with point defects separating
different boundary regions. Generically, the boundary lines may be gapped or gapless. The gapped cases are of
interest for us since they support point defects that have nontrivial topological properties.

We prove a classification of topologically distinct line defects of general Abelian topological states, extending the results of previous works\cite{bravyi1998,kapustin2010,kapustin2011,beigi2011,kitaev2012,wang2012,levin2013}.
In particular, it was proven recently in Ref. \onlinecite{levin2013} that a gapped boundary in an Abelian state
is determined by a ``Lagrangian subgroup" which consists of certain maximal subsets of topological quasi-particles that have trivial self and
mutual statistics. In this paper, we prove that \it every \rm such Lagrangian subgroup corresponds to
a topologically distinct gapped edge, and that Lagrangian subgroups therefore provide a classification of topologically distinct gapped edges.
Assuming that the notion of topological boundary conditions studied in Ref. \onlinecite{kapustin2011} is equivalent to gapped boundaries of
local Hamiltonians, this proves the classification that was conjectured in Ref. \onlinecite{kapustin2011}. Our proof is constructive,
in the sense that given an arbitrary Lagrangian subgroup $M$,
we show explicitly what \it local \rm operators to add to the edge theory to gap the edge in a way that corresponds to $M$.\footnote{
As this paper was being completed, we learned that this result has been independently derived by M. Levin and included
updated version of Ref. \onlinecite{levin2013}. The results of our paper build on the first version of Ref. \onlinecite{levin2013}.}

We further show that the nontrivial point defects on the boundary are then classified by the domain wall between gapped edges corresponding
to different Lagrangian subgroups. We compute the quantum dimension for generic point defects, which demonstrates
that the point defects are non-Abelian, and we develop an understanding of the zero modes that are topologically localized
to the point defects. These ``generalized parafermion'' zero modes vastly generalize the well-known Majorana fermion zero modes that
are currently under intense theoretical and experimental investigation.\cite{alicea2012review}
Although it is generally not possible to geometrically braid the point defects that live on the boundary,
we demonstrate that effective ``braiding" operations can be realized
in general by quasi-particle tunneling processes between pairs of defects.
Such braiding operations are topologically robust unitary transformations of the topologically degenerate states. We show
that they can always be mapped to the braiding of genons in a bilayer system.

We also studied the quantum phase transitions between different gapped boundary states realized on the same line defect.
Interestingly, the transition between two different types of boundary states $M$ and $M'$, corresponding to two different
Lagrangian subgroups, can be realized by nucleation of a periodic array of $M'$ regions in $M$. The domain walls between
$M$ and $M'$ regions define a periodic array of lattice defects, each of which supports non-Abelian zero modes. This
approach allows us to describe the quantum phase transition by a quantum spin chain that characterizes the coupling between
the topological zero modes. Alternatively, the spin chain can be formulated as a ``generalized parafermion chain,'' and
we expect that their phase transitions may be described by generalized parafermion conformal field theories.

We would like to further clarify the relation of our work with some previous works in the literature.
Gapped edges have been considered in recent years in several works. Ref. \onlinecite{beigi2011} constructed a set
of gapped edges for quantum double models, which are microscopic models of topologically ordered states
described at low energies by an emergent discrete gauge theory. These are restricted to time-reversal invariant
bosonic systems. It is not clear whether that construction
provides a complete classification of all possible gapped edges for those models.
Subsequently, Ref. \onlinecite{kitaev2012} developed a systematic microscopic analysis of gapped edges for a class of exactly soluble bosonic lattice models --
the Levin-Wen models\cite{levin2005} -- which pertain to both Abelian and non-Abelian states of time-reversal and parity symmetric
bosonic systems. Ref. \onlinecite{kapustin2011} studied ``topological boundary conditions'' of
Abelian Chern-Simons theory for bosonic systems, and conjectured that they are classified by Lagrangian subgroups. As pointed out
in Ref. \onlinecite{levin2013}, it is not clear whether topological boundary conditions are equivalent to gapped
boundaries of local Hamiltonians; Ref. \onlinecite{levin2013} further proved that the existence of a Lagrangian subgroup is a necessary and
sufficient condition for when an Abelian topological phase of bosons or fermions, realized by a local Hamiltonian,
can possibly admit a gapped edge. Ref. \onlinecite{fuchs2012} also studied topological boundary conditions in topological
quantum field theories (TQFTs) from a mathematical point of view, utilizing the framework of category theory.
However, we would like to emphasize that it is not clear whether the topological boundary conditions of TQFTs
that are classified in Ref. \onlinecite{fuchs2012} are equivalent to gapped edges of local Hamiltonians, which are the focus
of this paper and of Ref. \onlinecite{levin2013}.

We note that a portion of the results discussed in this paper were also reported by us in a recent shorter
treatment.\cite{barkeshli2013defect}

The rest of the paper is organized as follows. In Sec. \ref{TOsec}, we briefly review the formalism
for characterizing topological order, and the Abelian Chern-Simons theory framework for characterizing
all Abelian topological states. In Sec. \ref{foldingSec}, we introduce in more detail the notion of extrinsic line
and point defects in topological states, and discuss the mapping of all such defects to boundary
defects of topological states. In Sec. \ref{classificationSec}, we discuss gapped boundary defects
of topological states, and prove the classification of line defects in terms of Lagrangian subgroups
referred to above. In Sec. \ref{dwSec}, we study the topological properties of point defects as domain walls between
topologically distinct gapped edges. We derive a general formula for their quantum dimension, a
general understanding of the localized ``parafermion'' zero modes on the domain walls, and we discuss their
non-Abelian braiding statistics. In Sec. \ref{transSec}, we discuss the critical phenomena between
topologically distinct gapped edges; we show that this can be mapped onto the physics of a generalized
quantum spin chain, or, equivalently, a ``generalized parafermion'' chain. We conclude with a discussion in
Sec. \ref{discSec}.

\section{Characterization of Topological order}
\label{TOsec}

General topologically ordered states in 2+1 dimensions are characterized by the topological properties of
a set of topologically non-trivial quasiparticle excitations, $\{\gamma_i \}$, for $i = 1, \cdots, N_{qp}$,
where $N_{qp}$ is the number of quasiparticles. When two quasiparticles are observed from far away,
they in general behave like a superposition of single quasiparticle states. This is described by the
fusion rules $\gamma_i \times \gamma_j = \sum_k N_{ij}^k \gamma_k$. Secondly, when
two quasiparticles $\gamma_i$, $\gamma_j$ wind around each other, a phase $e^{i\theta_{ij}^k}$ is obtained,
which depends on the fusion channel $k$. $\theta_{ij}^k$ is referred to as the braid statistics of the
quasiparticles. When a particle is spinned around itself by $2\pi$, it generically gains a non-trivial
phase $e^{i\theta_i}$. $\theta_i = 0$ for bosons and $\pi$ for fermions. For a topological phase where the microscopic
degrees of freedom are all bosons, $\theta_i$ is topologically well-defined modulo $2\pi$.
In contrast, when the microscopic degrees of freedom also
contain fermions, then $\theta_i$ is topologically well-defined only modulo $\pi$. The braiding,
fusion rules, and spins must satisfy some consistency conditions, which we will not review here.\cite{preskillLectures,kitaev2006,zhwang2010}

A topological state is ``Abelian'' when quasiparticles at fixed locations do not induce additional
topological ground state degeneracies. This is equivalent to the condition that $N_{ij}^k =1 $ for
only one value of $k$, and $0$ otherwise.

A systematic description of all Abelian topological states is given by Abelian Chern-Simons
(CS) theory,\cite{wen04,wen1995} described by the Lagrangian density
\begin{align}
\label{CSLag}
\mathcal{L}_{CS} = \frac{1}{4\pi} K_{IJ}  \epsilon^{\mu \nu \lambda} a^I_\mu \partial_\nu a^J_\lambda ,
\end{align}
where $a^I$ for $I = 1, \cdots, \text{rank}(K)$ are compact $U(1)$ gauge fields,
$K$ is a non-singular, integer symmetric matrix, and $\mu$, $\nu$, $\lambda$ are $2+1$ dimensional space-time indices.
The topologically non-trivial quasiparticles are described by integer vectors $\b l$,
where two integer vectors $\b l$ and $\b l'$ describe topologically equivalent quasiparticles if
$\b l' = \b l + K \b \Lambda$, where $\b \Lambda$ is an integer vector. Therefore the integer lattice in $\text{rank}(K)$ dimensions,
modulo this equivalence relation, defines a discrete group consisting of the quasiparticles,
with the number of topologically distinct quasiparticles given by $|\text{Det } K|$. The exchange
statistics of a quasiparticle labelled by $\b l$ is given by $\theta_{\b l} = \pi \b l^T K^{-1} \b l $, and the mutual statistics of two
quasiparticles $\b l$, $\b l'$ is $\theta_{\b l \b l'} = 2\pi \b l^T K^{-1} \b l'$. $\theta_{\b l \b l'}$ is defined modulo $2\pi$,
while $\theta_{\b l}$ is defined modulo $2\pi$ for a topological phase of bosons, and modulo $\pi$ if the
microscopic Hamiltonian includes fermions. Vectors $K\b \Lambda$ describe local particles, which are always bosons or fermions. If all diagonal elements of $K$ are even integers (referred to as $K$ being even), then all local particles are bosons, and the theory describes a topological phase of bosons; otherwise
we say $K$ is odd, and the microscopic degrees of freedom must contain fermions (possibly in addition to bosons).

Different $K$-matrices can specify equivalent topological states if they have the same
quasiparticle content. For example, the transformation $K \rightarrow W^T K W$, for $W$
an integer matrix with $|\text{Det } W| = 1$, yields a different $K$-matrix, but describing the same
topological order. Alternatively, consider extending the $K$-matrix as follows:
\begin{align}
K' = \left(\begin{matrix} K & 0 \\ 0 & K_0 \end{matrix}\right)
\end{align}
where $K_0$ is an even-dimensional matrix with unit determinant and
zero signature (equal number of positive and negative eigenvalues).
Since $|\text{Det }K_0| = 1$, extending $K$ to $K'$ in this way
does not add any additional topological quasiparticles. Therefore $K'$ and
$K$ also describe the same topological order, as the group of quasiparticles and their
statistics is the same.

Eq. (\ref{CSLag}) possesses gapless edge states described by a $1+1$D chiral Luttinger liquid theory:\cite{wen04}
\begin{align}
\label{edgeL}
\mathcal{L}_{edge} = \frac{K_{IJ}}{4\pi} \partial_x \phi_I \partial_t \phi_J - V_{IJ} \partial_x \phi_I \partial_x \phi_J,
\end{align}
where $V_{IJ}$ is a positive-definite ``velocity'' matrix.
The number of left- and right-moving bosons, $n_L$ and $n_R$, are set by the number of positive
and negative eigenvalues of $K$, respectively.
The electron operators $\Psi_I$ and quasiparticle operators $\rchi_{\b l}$ on the edge are given by
\begin{align}
\Psi_I = e^{i K_{IJ} \phi_J}, \;\; \rchi_{\b l} = e^{i \b l^T \phi},
\end{align}
where $\b l$ is an integer vector describing the quasiparticles.
When $\Psi_I$ has integer scaling dimension, the ``electron'' is a boson, and if it is half-integer,
it is a fermion.

The Lagrangian (\ref{edgeL}) is gapless. When an Abelian topological state admits a gapped edge, it can be obtained by adding
additional backscattering terms in (\ref{edgeL}) to generate an energy gap in the edge theory.

\section{Line and Point Defects}
\label{foldingSec}

\begin{figure}
\centerline{
\includegraphics[width=3in]{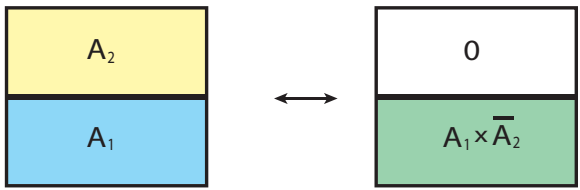}
}
\caption{\label{edgeFig} A line defect can be considered to be a domain wall between two kinds of topological phases,
$A_1$ and $A_2$. By folding one side over onto the other, this can be mapped to an edge between $A_1 \times \bar{A_2}$,
and the trivial gapped state, which we label ``0''. Under general conditions, the line defect will either host topologically
protected gapless edge states or be fully gapped.
}
\end{figure}

A general line defect in a topological state is a one-dimensional boundary between two topological states,
$A_1$ and $A_2$ (see Fig. \ref{edgeFig}). In some cases, such as when $A_1$ and $A_2$ have gapless edge states with
differing chiral central charges, the boundary possesses topologically protected gapless edge states.
In other cases, assuming certain criteria \cite{haldane1995,levin2013} that we review below are met, it is possible for
the boundary to be gapped. In this paper, we will only consider line defects that correspond to gapped boundaries.

The topological phases $A_1$ and $A_2$ do not necessarily have to be distinct: If
$A_1 = A_2 = A$, there can still be many different kinds of line defects. These correspond to
situations where quasiparticles are permuted amongst themselves as they cross the boundary, in a way which
preserves their topological quantum numbers.\cite{barkeshli2013genon,beigi2011,kitaev2012}
Such line defects are ``invisible,'' in the sense that braiding and fusion of quasiparticles on either side of the line
defect yields the same results.

In order to understand the properties of general boundaries, it is helpful to apply a
folding process, which has been employed previously in Ref. \onlinecite{beigi2011,kapustin2011},
and which we review here.  In order to understand the boundary between $A_1$ and $A_2$,
for concreteness we can consider $A_1$ and $A_2$ on a sphere, or plane, and then fold back
$A_2$ onto $A_1$, to obtain a boundary between the topological phase $A_1 \times \bar{A}_2$,
and the topologically trivial gapped phase, which we label ``0''. $\bar{A}_2$ denotes the parity-reversed
copy of $A_2$, which is necessary since the folding operation changes the parity of the state
that is being folded, because one of the directions is being reversed. Therefore, to study line defects,
it suffices to consider all possible boundaries between generic topological phases and the trivial
gapped phase.

\begin{figure}
\centerline{
\includegraphics[width=3in]{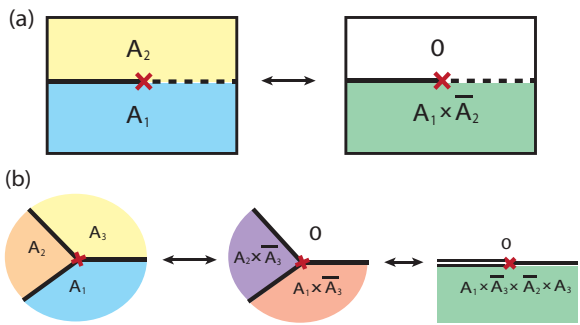}
}
\caption{\label{pointFig} (a) A domain wall between two different kinds of gapped edges separating
topological phases $A_1$ and $A_2$. By folding $A_2$ over, this can be mapped to a domain wall on the boundary separating
$A_1 \times \bar{A}_2$ and the trivial gapped state, ``0''. (b) A junction where multiple gapped edges meet is
also a possible type of point defect. On an infinite plane, by applying the folding trick multiple times, this can also be
mapped to a domain wall on the boundary separating a topological phase and $0$.
}
\end{figure}

Given the possibility of different kinds of gapped edges between topological phases, it is also possible to
have domain walls and junctions between them (see Fig. \ref{pointFig}). These point-like defects can
localize exotic topological zero modes, giving rise to topological ground state degeneracies, and projective
non-abelian statistics. In the special case where the line defects separate the same topological
phase on either side, the point defects are ``twist defects'' \cite{barkeshli2013genon} : As
a quasiparticle encircles the defect, it gets permuted by a symmetry of the topological quantum numbers.
Various examples of this have been studied previously in the literature\cite{bombin2010,barkeshli2012a,you2012,you2013,lindner2012,clarke2013,cheng2012}. In some of these physical realizations, such as those in Ref. \onlinecite{barkeshli2012a},
it is possible that the ground state energy of the system in the presence of the defects depends logarithmically on the
separation between them, as opposed to the more general linear energy cost for separating generic point defects.

Using the folding process, we can also understand point defects by mapping them to
domain walls between different gapped edges separating a topological phase and $0$ (see Fig. \ref{pointFig}).
In the case of junctions where multiple edges meet (Fig. \ref{pointFig}b), then we may apply the folding, in conjunction
with deformations of the location of the edges, several times in order to map the original configuration onto
a domain wall between different gapped edges separating a topological phase and $0$.
We note that this folding does not directly apply if the whole system is on a spatial manifold of arbitrary
topology, or for arbitrary configurations of point defects, but it is useful to understand the topological
behavior of each point defect by considering it in isolation on a plane, and then applying the folding process.

Therefore, in what follows we need only focus on gapped boundaries between a generic Abelian topological phase
and the trivial phase $0$, and domain walls between different such gapped boundaries.

\section{Classification of line defects}
\label{classificationSec}

\subsection{Review of null vectors and Lagrangian subgroups}
\label{nullVecSec}

As a starting point of our discussion, we review the concepts of null vectors and Lagrangian subgroups
discussed in Ref.  \onlinecite{haldane1995,levin2013}, which are the basic tools to obtain our new results. We consider
a generic Abelian topological phase, characterized by an Abelian CS theory with generic $K$-matrix.
The edge theory is described by (\ref{edgeL}). Backscattering terms can be added
on the edge, with the restriction that they be \it local \rm operators
on the edge. For systems involving fermions as the microscopic local degrees of freedom,
the backscattering terms must also conserve fermion number modulo $2$, known as the {\it fermion parity} symmetry.
Therefore the allowed backscattering terms on the edge are of the form
\begin{align}
\delta H_b = \sum_i \alpha_i(x) \cos(\b \Lambda_i^T K \phi + \theta_i(x) ),\label{eqn:backscattering}
\end{align}
where $\b \Lambda_i$ are integer vectors, $\b \Lambda_i^T K \b \Lambda_i$ is
even to ensure that the cosine terms are bosonic operators (\it ie \rm
have integer scaling dimension), and $\alpha_i(x)$ and $\theta_i(x)$ are spatially varying
functions.\footnote{For general choices of $\{\b \Lambda_i\}$, each backscattering term may also need to be multiplied by an
additional Klein factor, in order to ensure that they commute at separate points in space and
are therefore consistent with causality.}
When the number of left- and right- movers are unequal, $n_L \neq n_R$,
the edge states cannot be fully gapped. When $n_L = n_R = N$, \it ie \rm
there are an equal number of counterpropagating modes, it is possible but not guaranteed that
the edge be fully gapped, \it even in the absence of any symmetry. \rm
In fact, it has been shown that (\ref{edgeL}) can be fully gapped if and only if
there exist $N$ linearly independent vectors $\{\b \Lambda_i\}$ satisfying \cite{haldane1995,levin2013}:
\begin{align}
\label{nullCond}
\b \Lambda_i^T K \b \Lambda_j = 0.
\end{align}
This is a highly non-trivial constraint; for example, as discussed in Ref. \onlinecite{levin2013}, the $\nu=2/3$ FQH
edge, described by the $K$-matrix $K = \left(\begin{matrix} 1 & 0 \\ 0 & -3 \end{matrix} \right)$
does not admit such null vectors and therefore the edge cannot be gapped, \it even when particle
number conservation is broken \rm. On the other hand, the edge of the $\nu = 8/9$ FQH state described
by $K = \left(\begin{matrix} 1 & 0 \\ 0 & -9 \end{matrix} \right)$ can be gapped if particle number conservation
is broken.

The fact that such a set of null vectors $\{\b \Lambda_i\}$ can cause an energy gap on the edge
can be seen as follows. We perform a transformation $\phi = W\phi'$, such that
$\phi'$ has the commutation relations of a usual $N$-channel Luttinger liquid,
$[\phi'_i(x), \phi'_j(y)] = \pm \delta_{ij} i \pi sgn(x - y)$; $\phi'_i$ for $i = 1,\cdots, N$
can be chosen to be the left- movers, while for $i  =N+1, \cdots 2N$ they are the right-movers.
Under this condition, $\cos(\b \Lambda_i^T K \phi)$ becomes a conventional backscattering
term $\cos(\phi_i' \pm \phi_{i+N}')$.
It follows that the ground states can be characterized by the classical minima of the cosine terms:
$\b \Lambda_i^T K \phi = 2\pi n_i$, for $n_i \in \mathbb{Z}$.

The gapped boundary induced by the back-scattering terms (\ref{eqn:backscattering}) can be understood as
a one-dimensional ``condensate" of certain topological particles. The different components of the
integer-valued vector $K\b\Lambda_i$ may have common factors. Denote $K\b\Lambda_i=c_i\b m_i$ with
$c_i\in\mathbb{Z}$ and $\b m_i$ the minimal integer vector with no common factor in its components.
Since $\b \Lambda_i^T K \phi=2\pi n_i$ obtains a classical value on the edge, so does the quasiparticle operator
$e^{i\b m_i^T\phi}$, which satisfies $\left\langle e^{i\b m_i^T\phi}\right\rangle=e^{i2\pi n_i/c_i}$. Therefore the
quasiparticle $\b m_i$ is also condensed on the boundary line.

Taking this point of view, a gapped boundary can be generically viewed as a particle condensate. The condensed
quasiparticles form a subgroup $M$ of the group of all particles, with the group multiplication defined by particle
fusion. For the particle condensate to be defined consistently, the subgroup $M$ must satisfy the following conditions:
\begin{enumerate}
\item $e^{i \theta_{\b m\b m'}} = 1$ for all $\b m$, $\b m'$ $\in M$, and
\item $e^{i \theta_{\b l\b m}} \neq 1$ for at least one $\b m \in M$, if $\b l \notin M$.
\end{enumerate}
For bosonic states ($K$ even), we also have $e^{i \theta_{\b m}} = 1$ for all $\b m \in M$. The subgroup
$M$ has been referred to as a ``Lagrangian subgroup.''\cite{levin2013,kapustin2011,fuchs2012}

The first condition requires that every two particles in $M$ are mutually bosonic, so that they can be condensed simultaneously.
The second condition requires that all other quasiparticles not in $M$ are confined after the condensation of $M$.
Consequently the resulting state has no topologically non-trivial quasiparticle excitations that can propagate along the edge.

Following the discussion above, one can see that null vectors in back-scattering terms can be related to the condensation
of a Lagrangian subgroup on the edge, but the two are not obviously equivalent. Particles $\b m\in M$ in a Lagrangian
subgroup are not necessarily null vectors. Gapped edges that correspond to different Lagrangian subgroups $M$
are clearly topologically distinct; Ref. \onlinecite{levin2013} showed that every gapped edge corresponds to a
choice of $M$, and that every system with at least one Lagrangian subgroup has at least one type of gapped edge.
Here, we will strengthen this result by proving that \it every \rm Lagrangian subgroup $M$
corresponds to a gapped edge that condenses $M$. This shows that Lagrangian subgroups can {\it classify} gapped edges.

\begin{figure}
\centerline{
\includegraphics[width=2.5in]{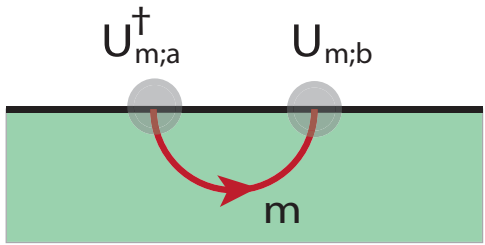}
}
\caption{\label{lineOps} Depiction of the process $W_{\b m}(\gamma)$, which creates
a quasiparticle $\b m$ at location $a$ on the edge with a local operator, the quasiparticle propagates
along a path $\gamma$ in the bulk, and is annihilated at $b$ with a local operator.
}
\end{figure}

Before presenting the proof of the classification in the next subsection, we discuss some other useful
properties of Lagrangian subgroups. The condensation of the quasiparticle set $M$ along the edge means
that a local operator can annihilate the quasiparticles in $M$ at the boundary, but
not in the bulk. As we will show in the following subsection, this
generally also implies that the operators in the edge theory
corresponding to quasiparticles in $M$ acquire a non-zero expectation value.
This condensation at the edge introduces a new process (Fig. \ref{lineOps}), whereby the system can start in the
ground state, a quasiparticle can be created at a point on the edge by a local operator, propagate
through the bulk, and then get annihilated by another local operator at a different point on the edge, no matter
how far apart the two points on the edge are.\cite{levin2013} We let this process be described by a quasiparticle
line operator $W_{\b m}(\gamma)$, where $\gamma$ is the path of the quasiparticle $\b m$. In the effective field theory,
\begin{align}
\label{lineOpDef}
W_{\b m}(\gamma) = U_{\b m;a} e^{i \b m^T \int_\gamma a\cdot dl} U_{\b m;b}^\dagger .
\end{align}
$e^{i \b m^T \int_\gamma a\cdot dl}$ is the Wilson line operator describing the propagation of the quasiparticle $\b m$
through the bulk, while $U_{\b ma}$ and $U_{\b mb}^\dagger$ are the local operators on the edge that annihilate/create
the quasiparticle $\b m$ at the points $a$, $b$, respectively. This process leaves the system in its ground state:
\begin{align}
W_{\b m}(\gamma) |\psi \rangle = |\psi \rangle,
\end{align}
where $|\psi\rangle$ is a ground state of the system.
These processes play a fundamental role in the proof that every gapped edge condenses a Lagrangian
subgroup. \cite{levin2013} As we will show, they similarly play a fundamental role in the analysis of
junctions between different gapped edges.

It is helpful to understand the role of the Lagrangian subgroup in cases where there is a
boundary between two topological states, $A_1$ and $A_2$. In the previous section, we discussed
the folding process, where we can consider the boundary between $A_1$ and $A_2$ as a
boundary between $A_1 \times \bar{A}_2$ and the trivial state. A gapped interface between $A_1$
and $A_2$ is folded to a gapped boundary of $A_1\times \bar{A}_2$. If the gapped boundary corresponds
to a Lagrangian subgroup $M$, the condensed particles $\b m \in M$ live in $A_1\times \bar{A}_2$. In
other words, $\b m$ is a pair of quasiparticles $\b m=({\b q}_1,{\bar{\b q}}_2)$ with ${\b q}_{1}$ and
${\bar{\b q}}_2$ a quasiparticle in $A_1$ and $\bar{A}_2$, respectively. If quasiparticle $({\b q}_1,{\bar{\b q}}_2)$
is condensed at the boundary, in the unfolded picture that means ${\b q}_1$ and ${\bar{\b q}}_2$ can be
brought to the boundary and annihilate each other. Consequently, ${\b q}_1$ can cross the boundary and
become ${\b q}_2$, the antiparticle of ${\bar{\b q}}_2$.
Therefore in this case each Lagrangian subgroup $M$ specifies a consistent set of transmission, reflection, and absorption processes that happens at the interface.

In the special case of twist defects,\cite{barkeshli2013genon} $A_1 = A_2 = A$. A Lagrangian subgroup of $A\times \bar{A}$ contains the pairs $({\b q}_1,\bar{\b q}_2)$ with ${\b q}_1, {\b q}_2\in A$. Therefore each Lagrangian subgroup defines a mapping ${\b q}_1\rightarrow {\b q}_2$ in theory $A$. The property that all $({\b q}_1,\bar{\b q}_2)$ in the Lagrangian subgroup are mutually bosonic is equivalent to the condition that the mapping ${\b q}_1\rightarrow {\b q}_2$ preserves the braiding and fusion rules of $A$. Therefore we correctly reproduced the known fact that twist defects are are in one-to-one correspondence with the symmetries of the
topological quantum numbers of $A$. \cite{barkeshli2013genon}

\subsection{From Lagrangian subgroups to backscattering terms}
\label{proofSec}

In this section, we sketch the proof that there is a one-to-one correspondence between Lagrangian subgroups $M$ and
sets of null vectors $\{\b \Lambda_i\}$ which can be used to gap the edges and condense the set
$M$ on the boundary.
An important condition for this proof to be valid is that it is allowed to couple the edge state to topologically trivial one-dimensional states.
This section expands the discussion in Ref. \onlinecite{barkeshli2013defect} recently presented by the authors.
Since our proof is constructive, our analysis provides a way to systematically construct the
local backscattering terms in the edge theory that condense any given Lagrangian subgroup $M$.

We will develop the argument for this in two steps. In the first step, we prove that a Lagrangian
subgroup defines a back-scattering term on the boundary if it is generated by a set of null
quasiparticles (defined below). In the second step, we prove that every Lagrangian subgroup can
be generated by a set of null quasiparticles, as long as it is possible to introduce purely
one-dimensional edge degrees of freedom which couple to the topological edge state.

\subsubsection{From null quasiparticles to back-scattering terms}
Consider a $2N \times 2N$ $K$-matrix with zero signature, and a Lagrangian subgroup $M$ generated by  $N$ linearly independent
$2N$-component integer vectors $\{\b m_i\}$. We assume that ${\b m}_i$ are null quasiparticles, which satisfy the condition
\begin{align}
\label{nullQP}
\b m_i^T K^{-1}\b  m_j = 0, \;\; i,j = 1,...,N.
\end{align}
Then, we define
\begin{align}
\label{lambdaDef}
\b \Lambda_i  = c_i K^{-1} \b m_i,
\end{align}
where $c_i \in \mathbb{Z}$ is the minimal integer such that $\b \Lambda_i$ is an integer vector.
The $\{\b \Lambda_i\}$ defined this way satisfy (\ref{nullCond}), and therefore
\begin{align}
\label{backSc}
\delta H = g \sum_{i=1}^N \cos(\b \Lambda_i^T K \phi)
\end{align}
can generate an energy gap in the edge states, by pinning the argument of the cosine terms:
\begin{align}
\b \Lambda_i^T K \phi = c_i \b m_i^T \phi = 2\pi n_i,
\end{align}
where $n_i$ is an integer. This directly implies that
\begin{align}
\label{qpCond}
\langle e^{i \b m_i^T \phi} \rangle = e^{i 2\pi n_i/c_i} \neq 0.
\end{align}
Therefore, the edges are gapped and the Lagrangian subgroup $M$ is condensed on the boundary.

\subsubsection{From general Lagrangian subgroups to null quasiparticles}

Now, we need to prove that every Lagrangian subgroup $M$ can be represented by $N$
linearly independent vectors $\{\b m_i\}$ which satisfy (\ref{nullQP}). Naively, for a given
$K$-matrix this is not true. For example, consider $K = \left(\begin{matrix} 0 & 4 \\ 4 & 0 \end{matrix} \right)$,
which describes $Z_4$ topological order. This has a Lagrangian subgroup generated by
$\b m_1^T = (2,0)$, $\b m_2^T = (0,2)$ which does not satisfy (\ref{nullQP}).
It turns out the statement that we will actually need is the following.

\bf{Lemma: } \rm Suppose we have a set of vectors $\{\b m_i\}$, $i = 1,..., N_M$,
(where $N_M$ is not necessarily equal to $N$), which generate a Lagrangian subgroup:
\it ie \rm $\b m_i^T K^{-1} \b m_j \in \mathbb{Z}$, $\b m_i^T K^{-1} \b l \notin \mathbb{Z}$
$\forall \b l \notin M$, and $\b m_i^T K^{-1} \b m_i$ is even for $K$ even.
Then, there exists a $K'$ which is topologically equivalent to $K$, such that $dim(K') = 2N'$,
and a set of $N'$-component vectors, $\{\b m_i'\}$, for $i=1,...,N'$, which satisfy
$\b m_i'^T K'^{-1} \b m_j' = 0$, and which generate the same Lagrangian subgroup $M$.

The proof of this is somewhat technical and will be presented in the appendix. The
main idea is that one can define $K'$ of the form
\begin{eqnarray}
K'=\left(\begin{array}{ccc}K&0&0\\0&0&\mathbb{I}\\0&\mathbb{I}&0\end{array}\right)\text{~or}
\left(\begin{array}{ccc}K&0&0\\0&\mathbb{I}&0\\0&0&-\mathbb{I}\end{array}\right)
\end{eqnarray}
with $\mathbb{I}$ an $N\times N$ identity matrix. The two forms should be applied to the $K$'s describing a boson theory or a fermion theory,
respectively. Since $|\text{Det }K'|=|\text{Det }K|$, the new blocks do not introduce any new particle types, and $K'$ and $K$ are topologically
equivalent. Every generator ${\b m}_i$ of the Lagrangian subgroup is mapped to a higher dimensional vector $\{\b m_i'\}$ by simply expanding
the $K$ matrix to include additional counterpropagating topologically trivial edge states. Although the added topologically trivial degrees of
freedom do not change the topological properties of quasiparticle ${\b m}_i$, it can change the inner product of ${\b m}_i'$. A suitable choice
can always be made to satisfy $\b m_i'^T K'^{-1} \b m_j' = 0$. In the next subsection, we will discuss a number of explicit examples of this.

The above lemma, taken together with step 1 above, prove that every Lagrangian subgroup $M$ of an Abelian
topological phase corresponds to a gapped edge where $M$ is condensed.
Therefore, the Lagrangian subgroups provide a topological classification of gapped edges.
Edges corresponding to different Lagrangian subgroups clearly cannot be adiabatically connected
to each other without closing the energy gap in the edge states.

\subsection{Examples}
\label{gappedEdgeEx}

Let us begin with a simple example. Consider two independent time-reversed copies of a
$1/m$-Laughlin FQH state, described by the $K$-matrix $K = \left( \begin{matrix} m & 0 \\ 0 & -m \end{matrix}\right)$.
The edge of this state can be terminated by a charge-conserving backscattering term,
\begin{align}
H_{b} = \frac{g}{2}\left( \Psi_{eL}^\dagger \Psi_{eR} + H.c. \right)  = g \cos( m(\phi_L + \phi_R))
\end{align}
or by superconductivity:
\begin{align}
H_{sc} = \frac{g}{2}\left( \Psi_{eL}^\dagger \Psi_{eR}^\dagger + H.c.\right) = g \cos(m (\phi_L - \phi_R)).
\end{align}
In the first case, the Lagrangian subgroup generated by $(1,1)$ is condensed on the edge;
in the second case, the Lagrangian subgroup generated by $(1,-1)$ is condensed on the edge.

Now let us consider a more non-trivial example that illustrates the necessity of the Lemma
introduced in the previous subsection. Consider $K = \left(\begin{matrix} 9 & 0 \\ 0 & -9 \end{matrix}\right)$.
This has a Lagrangian subgroup generated by $\b m_1^T = (3,0)$ and $\b m_2^T = (0,3)$. Now we
define
\begin{align}
K' = \left(\begin{matrix} K & 0 & 0\\ 0 & 1 & 0 \\ 0 & 0 & -1\end{matrix} \right),
\end{align}
and $\b m_1'^T = (3,0,0,1)$, $\b m_2'^T = (0,3,1,0)$. Again, in the absence of any symmetries,
$K'$ is topologically equivalent to $K$ and $\b m_i'^T K'^{-1} \b m_j' = 0$.
Thus the backscattering terms associated with
this Lagrangian subgroup are given by $\sum_{i=1}^2 \cos( \b \Lambda_i^T K' \phi)$, with
$\b \Lambda_i = 3 K^{-1} \b m_i'$.

Another example is given by the mutual Chern-Simons theory describing the $Z_4$ toric code model\cite{kitaev2003}.
Let $K =  \left(\begin{matrix} 0 & 4 \\ 4 & 0 \end{matrix} \right)$, and
$\b m_1^T = (2,0)$, $\b m_2^T = (0,2)$. We define
\begin{align}
K' = \left(\begin{matrix} K & 0 & 0 \\ 0 & 0 & 1 \\ 0 & 1 & 0  \end{matrix} \right),
\end{align}
and $\b m_1'^T = (2,0,0,1)$, $\b m_2'^T = (0,2,-1,0)$. Here, $K'$ is topologically equivalent
to $K$, and $\b m_i'^T K'^{-1} \b m_j' = 0$. Now the backscattering terms associated with
this Lagrangian subgroup are given by $\sum_{i=1}^2 \cos( \b \Lambda_i^T K' \phi)$, with
$\b \Lambda_i = 2 K^{-1} \b m_i'$.

Let us now consider more generally the $Z_N$ toric code model, described by
$K = \left( \begin{matrix} 0 & N \\ N & 0 \end{matrix} \right)$.
For any set of integers $r,t$ such that $rt = N$, there is a Lagrangian subgroup generated by the quasiparticles
$(r,0)$ and $(0,t)$.
In other words, every distinct divisor of $N$ yields a different Lagrangian subgroup. The number of
Lagrangian subgroups is therefore equal to the number of divisors of $N$. For example when $N$ is prime,
there are two Lagrangian subgroups, corresponding to whether the ``electric''
particles $(1,0)$ are condensed, or whether the ``magnetic'' ones, $(0,1)$ are condensed.\cite{bravyi1998}
When $N = \prod_i p_i^{s_i}$ where the $p_i$ are all distinct prime numbers and $s_i$ are their
multiplicities, there are $\prod_i (1+s_i)$ Lagrangian subgroups and therefore $\prod_i (1+s_i)$
topologically distinct boundaries.

\begin{figure}[tb]
\centerline{
\includegraphics[width=3.0in]{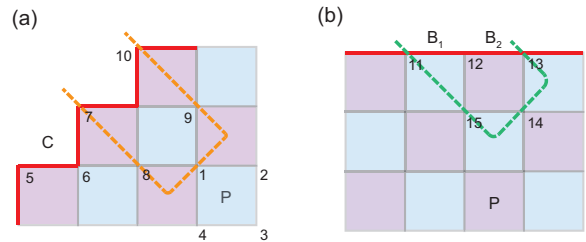}
}
\caption{(a) Lattice model of $Z_N$ toric code with gapped boundary (red line) that corresponds to the Lagrangian subgroup generated by $(1,0)$ and $(0,N)$. The Wilson line (the orange dash line) shows that the quasi-particle $(1,0)$ and its anti-particle $(-1,0)$ can be created together from the vacuum and annihilated at different locations of the boundary. (b)
The $Z_N$ toric code with gapped boundary (red line) that corresponds to the Lagrangian subgroup generated by $(r,0)$ and $(0,t)$. The Wilson line (the green dash line) shows that the quasi-particle $(0,t)$ and its anti-particle $(0,-t)$ can be created together from the vacuum and annihilated at different locations of the boundary.
\label{tcEdge}
}
\end{figure}

In the following discussion, we will provide a microscopic lattice model construction of these different gapped
edges of the $Z_N$ toric code. We will use a specific construction of this phase, called the $Z_N$ plaquette model. \cite{wen2003,schulz2012,you2012}
The degrees of freedom consist of $N$ states on each of the sites of a square lattice.
The Hamiltonian with the gapped edge contains two terms:
\begin{align}
H_{\text{Total}}=H_{\text{Bulk}}+H_{\text{Edge}},
\end{align}
where
\begin{align}
H_{\text{Bulk}}=-\sum_{p} (\mathcal{O}_p+ h.c ).
\end{align}
The sum is over all plaquettes of the square lattice, and the plaquette operator
$\mathcal{O}_p$, as shown in Fig. \ref{tcEdge} (a), is defined as
\begin{align}
\mathcal{O}_p=T_1 U_2 T^\dag_3 U^\dag_4,
\end{align}
where $T_i$ and $U_i$ are $N\times N$ matrices satisfying $U_i T_i =T_i U_i e^{i \frac{2\pi}{N}}$, $T_i^N=U_i^N=1$.
The $N$ states at each site form an $N$-dimensional irreducible representation of this algebra.
Since all the plaquette operators $\mathcal{O}_p$ commute with each other, the ground state is the
common eigenvector of all $\mathcal{O}_p$'s with the real part of the eigenvalue maximized.
If the real part of the eigenvalue of $\mathcal{O}_p$ on a purple (blue) plaquette is not maximized,
the state contains a electric (magnetic) quasiparticle at that plaquette.
The choice of gapped edge is set by the nature of $H_{\text{Edge}}$. For a gapped edge that
corresponds to the Lagrangian subgroup generated by $(1,0)$ and $(0,N)$, we consider
the physical edge in Fig. \ref{tcEdge} (a) indicated by the red line, with
\begin{align}
H_{\text{Edge}}=-\sum_{C} (\mathcal{V}_C+ H.c),
\end{align}
which is a sum over all "corner operators" on the edge (see Fig. \ref{tcEdge} (a)). The corner operator
at the corner $C$ is defined as
\begin{align}
\mathcal{V}_C=U_7 T^\dag_6 U^\dag_5.
\end{align}
With this definition, all terms in $H_{\text{Total}}=H_{\text{Bulk}}+H_{\text{Edge}}$ commute with each other. The ground state
will be a common eigenvector of all $\mathcal{O}_p$'s and $\mathcal{V}_C$'s with the real parts of all the eigenvalues
maximized. By counting the number of constraints from this and the total number of degrees of freedom, we find there
are only a finite number of ground states with gapped excitations, which implies a gapped edge. Now we want to show
that the edge indeed corresponds to the Lagrangian subgroup generated by $(1,0)$ and $(0,N)$. We can consider the
process in which a electric particle-antiparticle pair is created in the bulk and then annihilated at different locations
on the edge. An example of this process is described by the Wilson line operator indicated by the orange dashed line Fig. \ref{tcEdge} (a):
\begin{align}
W_e=U_7 U_8 T^\dag_1 U^\dag_9 U^\dag_{10}.
\end{align}
Notice that $W_e$ can be written as a product of plaquette operators and corner operators. Thus, $W_e$ leaves the
ground state invariant, which means that the electric particle is condensed on the edge.
Moreover, in this construction, $(0,N)$ is a trivial particle, and therefore is already ``condensed'' on the edge. Thus the
model we write down here describes the $Z_N$ toric code with a gapped edge corresponding to the Lagrangian
subgroup generated by $(1,0)$ and $(0,N)$.

More generally, for the gapped edge that corresponds to the Lagrangian subgroup generated by $(r,0)$ and $(0,t)$, we consider
the physical edge in Fig. \ref{tcEdge} (b) indicated by the red line, with
\begin{align}
H_{\text{Edge}}=-\sum_{B} (\mathcal{R}_B+ h.c),
\end{align}
which is a sum over all "bond operators" on the edge as shown in Fig. \ref{tcEdge} (b). If the bond
is the edge of a magnetic plaquette, say bond $B_1$, we define
\begin{align}
\mathcal{R}_{B_1}= (T^\dag_{12} U^\dag_{11})^t,
\end{align}
while the bond operator on the edge of an electric plaquette, say bond $B_2$, is defined as
\begin{align}
\mathcal{R}_{B_2}= (T^\dag_{13} U^\dag_{12})^r.
\end{align}
Again, all terms in $H_{\text{Total}}=H_{\text{Bulk}}+H_{\text{Edge}}$ commute with each other.The ground state is a
common eigenvector of all $\mathcal{O}_p$'s and $\mathcal{R}_B$'s with the real parts of all the eigenvalues
maximized. By a similar analysis as above, the Hamiltonian $H_{\text{Total}}$ produces a gapped edge. Now, we want to
show that the particles $(r,0)$ and $(0, t)$ are condensed on this edge. We consider the process in which the pair
$(0,t)$ and $(0,-t)$ is created from the vacuum, and then annihilated at different locations on the edge.
This process can, for example, be described by the Wilson line operator $W_{tm}$ denoted in Fig. \ref{tcEdge}(b) by the green dashed-line:
\begin{align}
W_{tm}=(U_{11} U_{15} T^\dag_{14} U^\dag_{13} )^t.
\end{align}
Notice that $W_{tm}$ can be written as a product of plaquette operators and bond operators. Thus, $W_{tm}$
leaves the ground state invariant, which means that the particle $(0,t)$ is condensed on the edge of
the ground state. A parallel analysis can be performed for $(r,0)$ particle to show that it also condenses at the edge.
Therefore, this lattice construction corresponds to the Lagrangian subgroup generated by $(r,0)$ and $(0,t)$.

\section{Classification of point defects}
\label{dwSec}

Now let us consider junctions where various gapped edges meet at a point. As shown in Fig. \ref{pointFig},
using the folding process, on an infinite plane such junctions can always be mapped to domain walls
between two different gapped edges. Therefore, here we need only to focus on domain walls
between two different gapped edges in order to understand the essential topological properties
of generic point defects.

Thus, consider two kinds of gapped edges associated with two different
Lagrangian subgroups $M$ and $M'$. In order to understand basic properties such as
quantum dimension, zero modes, and non-abelian statistics, we will consider the system
on the disk geometry, with $2n$ well-separated domain walls separating the two gapped edges. We will refer to the
edges where a Lagrangian subgroup $M$ is condensed as an $M$-edge,
and similarly for edges where $M'$ is condensed.

\subsection{Topological degeneracies: Quantum Dimension}
\label{qdimSec}

\begin{figure}
\centerline{
\includegraphics[width=2.7in]{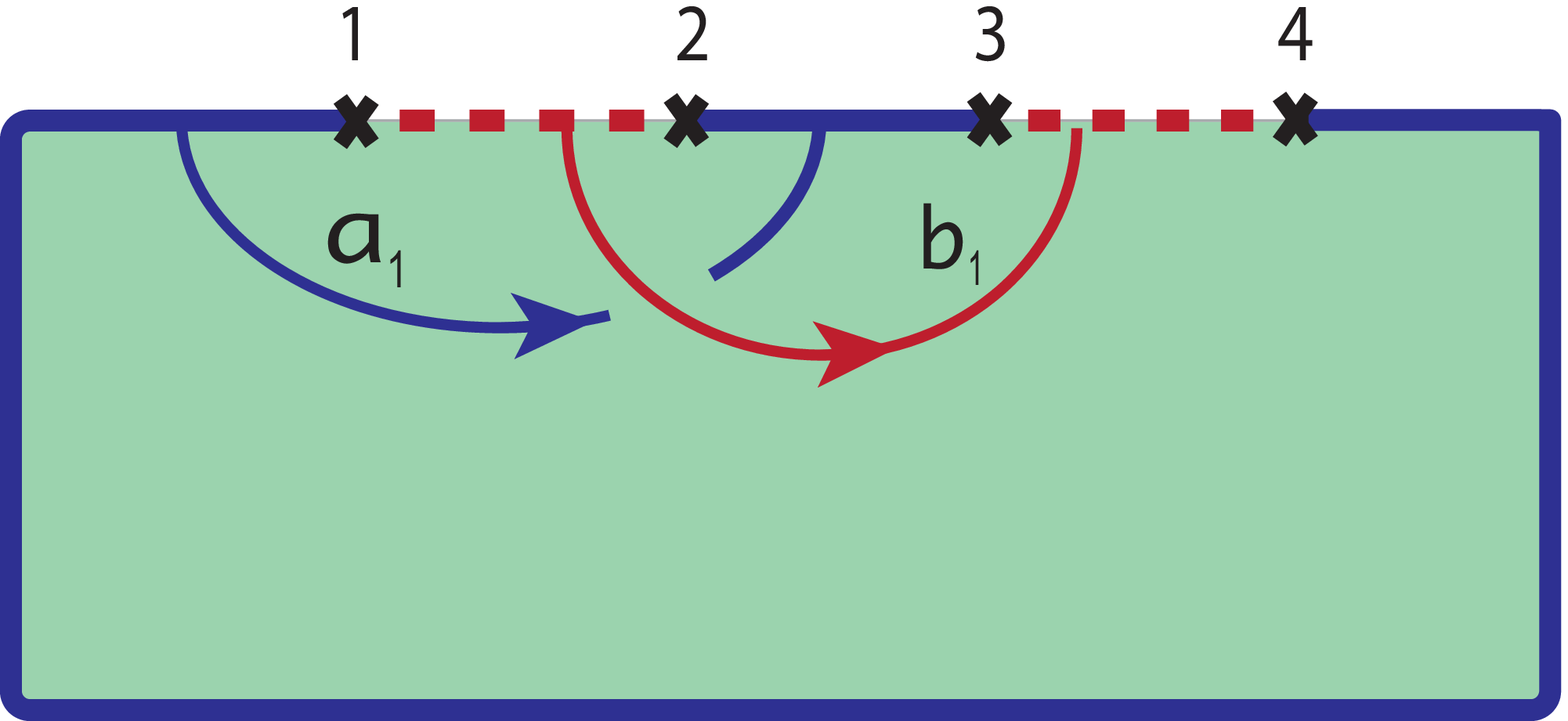}
}
\centerline{
\includegraphics[width=2.7in]{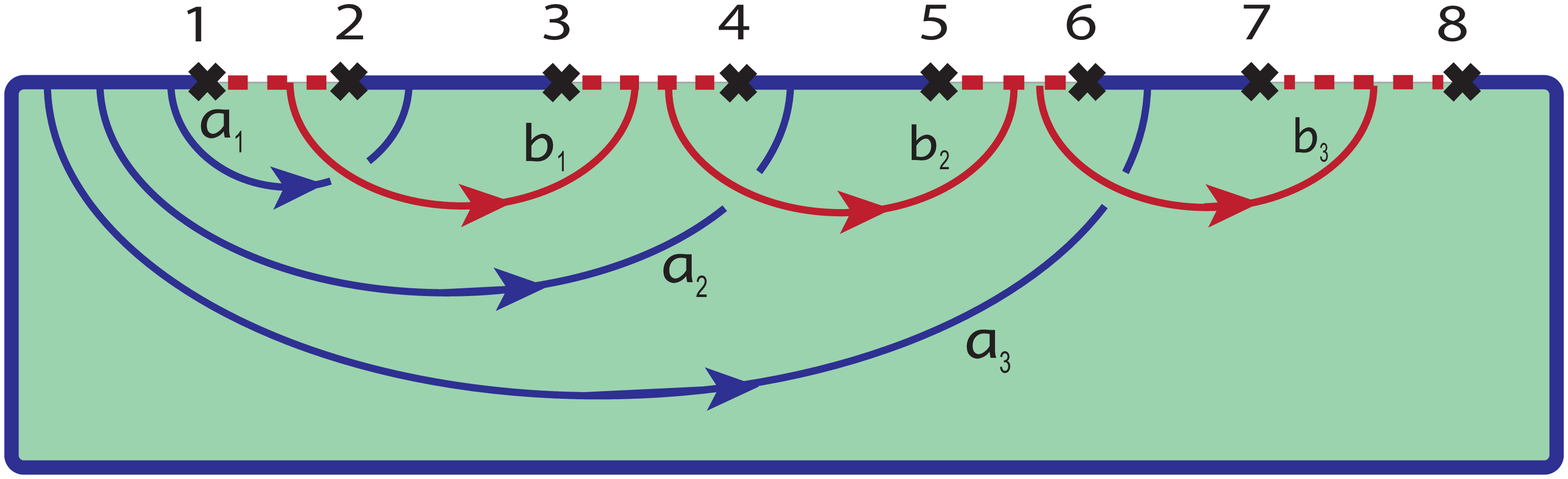}
}
\caption{\label{lineAlg} (Upper panel) A topological phase on a disk topology, with the trivial gapped state outside of the
disk. The solid blue (dashed red) lines on the edge correspond to an $M$ ($M'$)-edge. The black crosses indicate
the domain walls between the two kinds of gapped edges. The existence of the domain walls implies the possibility
of non-commuting Wilson line operators associated with quasiparticles of the two Lagrangian subgroups $M$
and $M'$. (Lower panel) The existence of many domain walls between gapped edges corresponding to Lagrangian subgroups
$M$ and $M'$ implies the possibility of many non-commuting Wilson line operators associated with quasiparticles of $M$ and $M'$.
A convenient set of paths, $\{a_i\}$ and $\{b_i\}$, is shown, which shows that for $n$ pairs of domain walls, there
are $n-1$ copies of the algebra (\ref{magAlg2}).
}
\end{figure}

The essential feature of the existence of domain walls between multiple gapped edges
is the introduction of novel line operators, $W_{\b m}(a_{i})$ and
$W_{\b m'}(b_i)$, where $\b m \in M$ and $\b m' \in M'$, with paths $a_{i}$
and $b_{i}$ that can intersect only once (see Fig. \ref{lineAlg}). The definition of these line operators, for the case of a
single gapped edge, was discussed in Sec. \ref{nullVecSec} and depicted in Fig. \ref{lineOps}.

Since $\b m$ and $\b m'$ have fractional mutual statistics, these operators do not commute with each other:
\begin{align}
\label{magAlg}
W_{\b m}(a_i) W_{\b m'}(b_j) = W_{\b m'}(b_j) W_{\b m}(a_i) e^{\delta_{ij} 2\pi i \b m^T K^{-1} \b m'},
\end{align}
where $a_i$ and $b_i$ are the paths shown in Fig. \ref{lineAlg} (b).
Since these line operators leave the system in its ground state subspace, the ground states must
form a representation of this algebra. The dimension of the smallest irreducible representation
of this algebra is generally larger than one, from which we can conclude that the domain walls
must introduce topological ground state degeneracies into the system.

These Wilson line operators are generalizations of the Wilson loop operators in the presence of twist
defects \cite{barkeshli2012a,barkeshli2013genon,barkeshli2013}, where the defects
introduce novel non-contractible loops that lead to a non-trivial loop algebra and therefore a topological
ground state degeneracy.

In what follows, we will provide a general formula for the ground state degeneracy, which allows us to
obtain the quantum dimension, or effective number of degrees of freedom, of each domain wall.

\subsubsection{Calculation of quantum dimension}

In the case where we have one pair of point defects, there is no non-trivial line algebra induced by the defects, and therefore the defects do not induce any topological degeneracy. Therefore we will begin with the case where we have two pairs of point defects (Fig. \ref{lineAlg}).

In order to compute the topological ground state degeneracy, we must find the dimension of
the smallest irreducible representation of (\ref{magAlg}). To do this, it is convenient to
first write the quasiparticles in terms of the generators of the Lagrangian subgroups $M$ and $M'$.
That is,
\begin{align}
\b m = \sum_{i=1}^{N} q_i \b m_i, \;\; \b m' = \sum_{i=1}^{N}q_i' \b m'_i,
\end{align}
where $q_i$ and $q_i'$ are integers, and $\{\b m_i\}$, $\{\b m'_i\}$, for $i=1,...,N$
are the generators of $M$ and $M'$, respectively. Here we use the $N$ null vectors
described in Sec. \ref{proofSec} for the generators of the Lagrangian subgroups, where $dim(K) = 2N$.
For simplicity we relabel the line operators
\begin{align}
A_{\vec{q}} = W_{\sum_i q_i \b m_i}(a_1), \;\;
B_{\vec{q'}} = W_{\sum_i q_i' \b m'_i}(b_1),
\end{align}
where $\vec{q}$ and $\vec{q'}$ above are $N$- component integer vectors.

In this notation, the line algebra is
\begin{align}
\label{magAlg2}
A_{\vec{q}} B_{\vec{q'}}= B_{\vec{q'}} A_{\vec{q}} e^{2\pi i \vec{q}^T R \vec{q'}},
\end{align}
where $R$ is an $N \times N$ matrix:
\begin{align}
R_{ij} = \b m_i^T K^{-1} \b m'_j.
\end{align}
The $A$'s and $B$'s
all commute with each other, and
\begin{align}
\label{fusion}
A_{\vec{q}} A_{\vec{q'}} = A_{\vec{q} + \vec{q'}}, \;\;
B_{\vec{q}} B_{\vec{q'}}= B_{\vec{q} + \vec{q'}}.
\end{align}
Note also that
\begin{align}
\label{cyclic}
A_{\vec q} &= 1, \;\;\; \text{ if } R^T \vec{q} \in \mathbb{Z}^{N},
\nonumber \\
B_{\vec{q'}} &=1, \;\;\; \text{ if } R \vec{q'} \in \mathbb{Z}^{N}.
\end{align}
This is because such $A_{\vec{q}}$ and $B_{\vec{q'}}$ will commute with all operators in the algebra,
and can therefore be represented as the identity in the ground state subspace.

The smallest irreducible representation of this algebra can be
obtained by diagonalizing one set, such as $A_{\vec q}$, and having $B_{\vec{q'}}$
act as the ladder operators:
\begin{align}
A_{\vec q} |\vec{\alpha} \rangle = e^{2\pi i \vec{q} \cdot \vec{\alpha}} |\vec{\alpha} \rangle,
\;\;
B_{\vec{q'}} |\vec{\alpha} \rangle = |\vec{\alpha} + R \vec{q'}\rangle
\end{align}
where $\vec{\alpha}$ is an $N$-component rational-valued vector. The fact that the eigenvalues
must be phases follows from (\ref{fusion}), and the fact that the $\vec{\alpha}$ are rational-valued
follows from (\ref{cyclic}). (\ref{cyclic}) also implies
\begin{align}
\label{equiv}
|\vec{\alpha} \rangle = |\vec{\alpha} + R \vec{q'}\rangle , \;\;\; \text{ if } R \vec{q'} \in \mathbb{Z}^{N}.
\end{align}
Therefore, the number of ground states can be obtained by counting the number of independent
possible values of $|\vec{\alpha}\rangle$. The ladder operators $B_{\vec{q}}$ define states associated with the
lattice $R \mathbb{Z}^{N}$, subject to the equivalence (\ref{equiv}). Therefore, $R$ can be viewed
as generating a lattice of rational-valued vectors, where each state corresponds to a point
on the lattice, and two states are equivalent if they differ by an integer-valued vector.

The number of such states can be computed as follows.
We consider the lattice
\begin{align}
\label{RtildeDef}
\Gamma = \{R \v \Lambda' + \v \Lambda: \v \Lambda'\in \mathbb{Z}^{N}, \v \Lambda \in \mathbb{Z}^{N}\} .
\end{align}
$\Gamma$ is an $N$-dimensional lattice, which can be generated by a
matrix:
\begin{align}
\label{RtildeDef2}
\Gamma = \tilde{R} \mathbb{Z}^{N}.
\end{align}
Now observe that $\mathbb{Z}^{N}$ is itself a sublattice of $\Gamma$. Therefore, the unit cell of $\mathbb{Z}^{N}$, which
has unit volume, must contain an integer number $D$ of unit cells of $\Gamma$. This implies that
the volume of each unit cell of $\Gamma$ is $1/D$: $|\text{Det }\tilde{R}| = 1/D$. Each unit cell can be associated with
one state, and the inequivalent states all exist inside the unit sublattice of $\mathbb{Z}^{N}$. Thus
there are $D$ states, so the dimension of the smallest irreducible representation of (\ref{magAlg}) is $D$.
In Sec. \ref{qdimExamples}, we will provide a number of concrete examples of this calculation.

In the case where we have $2n$ domain walls, there are $n-1$ independent copies of the above algebra (see Fig. \ref{lineAlg} (b)), and therefore
the ground state degeneracy is $D^{n-1}$. This implies that the quantum dimension of
each defect is
\begin{align}
\label{dbulk}
d = \sqrt{D}=\left(|\text{Det }\tilde{R} |\right)^{-1/2}.
\end{align}

We note that the topological degeneracy studied above is exact in the limit that the defects are infinitely far apart.
If the defects are separated by a finite distance $\ell$, then these topologically degenerate states will obtain
an energy splitting proportional to $e^{-\ell/\xi}$, where $\xi$ is the correlation length of either the edge states
or the bulk, depending on which one is larger. This is due to the fact that a finite separation of the defects
allows the instanton processes associated with the Wilson line operators to appear in the Hamiltonian and thus to
split the energy of the degenerate states. The amplitude for these terms decays exponentially as $e^{-\ell/\xi}$,
due to the energy gap of the system. This is similar to the case of topological degeneracies of topological states on
closed surfaces, where the finite system size induces exponentially small splittings among the degenerate ground states.\cite{wen1989}
This mechanism for inducing an energy gap among the degenerate ground states will be important for the
discussions of Sec. \ref{braidSec} and \ref{transSec}.

\subsubsection{General discussion of quantum dimension}
\label{qdimDiscSec}

The quantum dimension and associated ground state degeneracy calculated above is topologically
robust and derived from the fractional statistics of the bulk quasiparticles. Therefore it does not include
any possible additional topologically degeneracies that may arise from purely one-dimensional physics.
In the absence of any symmetries, it has been proven that bosons in one dimension cannot give rise to
any topological degeneracies \cite{chen2011}. Therefore, the result above fully
captures the quantum dimension for domain walls in bosonic systems.

In contrast, fermionic systems in one dimension, in the absence of any symmetry, have a $Z_2$ topological
classification \cite{kitaev2001,fidkowski2011,turner2011}. There can be domain walls between different
gapped 1D fermionic systems that localize Majorana fermion zero modes, which have a quantum dimension
of $\sqrt{2}$. These Majorana fermion zero modes are protected by fermion parity symmetry: If the system
is coupled to a gapless reservoir of fermions, the fermion parity symmetry
of the edge system will be broken, and the topological degeneracy associated with the Majorana modes
will be split. Therefore, for topological states that include at least one fermion species, the quantum
dimension of the defect is
\begin{align}
d_{defect} = d_{1D} d_{bulk},
\end{align}
where $d_{bulk}$ is the Wilson line algebra contribution of (\ref{dbulk}), and $d_{1D}$ is either $\sqrt{2}$ or $1$,
depending on whether the purely one-dimensional fermion physics has an extra Majorana zero mode, protected by fermion parity.
$d_{1D}$ is independent of the Lagrangian subgroup, and depends more precisely on the backscattering
Hamiltonian of the edge theory. It is possible in principle to compute $d_{1D}$, although
this requires information beyond just the Lagrangian subgroups $M$ and $M'$, and requires knowledge of the
precise Hamiltonian which is generating the energy gap on the edge.

We conclude that the Lagrangian subgroups can fully classify gapped edges only ``modulo one-dimensional physics.''
For bosonic systems, the Lagrangian subgroups are expected to provide a full classification, while for fermionic systems, it is possible
that two gapped edges correspond to the same Lagrangian subgroup, but cannot be adiabatically connected to each
other. If we define the equivalence by allowing arbitrary one-dimensional degrees of freedom to be added to the edges,
then of course the Lagrangian subgroups provide a full classification even for fermionic systems.

We note that here, we have computed the topological degeneracy by studying the representations of the line operators
that begin and end on the edge. This is a generalization to boundary domain walls of the loop algebra approach to computing
the quantum dimension of twist defects, used in Ref. \onlinecite{barkeshli2013genon}.
An equivalent way to compute the topological degeneracy is directly within
the edge theory, by studying the ground states of the chiral Luttinger liquid theory (\ref{edgeL}), with different
kinds of backscattering terms of the form (\ref{backSc}). Such an analysis was presented in special cases in
Ref. \onlinecite{barkeshli2012a,barkeshli2013genon,lindner2012,clarke2013,cheng2012}.  The main idea can be understood simply as follows.
The edge Hamiltonian can be written as
\begin{align}
\delta H_{edge} =
g\sum_{i = 1}^N \left\{ \begin{array}{ccc}
\cos( c_i \b m_i^T \phi ) & \text{ if } & x \in M \text{ region} \\
\cos(c_i' \b m_i'^T\phi) & \text{ if }& x \in M' \text{ region}\\
\end{array} \right.
\end{align}
Recall that $\{\b m_i\}$ and $\{\b m_i'\}$ are the $N$ generators of the $2N$-dimensional matrix $K$,
and $c_i$, $c_i'$ are the smallest integers such that $c_i K^{-1} \b m_i$, $c_i' K^{-1} \b m_i'$ are integer vectors,
as explained in Sec. \ref{proofSec}. The cosine terms in both domains cannot simultaneously acquire
their classical minimum values. From (\ref{backSc})-(\ref{qpCond}), we see this would require
$\langle e^{i \b m^T \phi (x) } \rangle \neq 0$ for $x$ in the $M$-gapped region, and
$\langle e^{i \b m'^T \phi (x) } \rangle \neq 0$ for $x$ in the $M'$-gapped region, which is not possible in general
because these two operators generically do not commute at different points in space,
and therefore they cannot be simultaneously diagonalized. Picking only the operators in one
region to be fully diagonalized then leads to a topological degeneracy that grows exponentially
with the number of domain walls. Such an understanding is useful for more detailed computations using the edge theory.

\subsubsection{Examples}
\label{qdimExamples}

Here we will review some examples, taken from previous studies\cite{lindner2012,clarke2013,cheng2012,
barkeshli2010,barkeshli2012a,barkeshli2013genon,bombin2010,you2012,you2013}, using the general framework developed here
in terms of the Wilson line algebra.

First, let us consider the proposals of Ref. \onlinecite{lindner2012,clarke2013,cheng2012}, which consider a FQH state with $K$-matrix
\begin{align}
K = \left(\begin{matrix} N & 0 \\ 0 & -N \end{matrix} \right).
\end{align}
The two Lagrangian subgroups $M$ and $M'$
considered are generated by $\b m_1 = (1, 1)$ and $\b m_1'=(1, -1)$, respectively. When these vectors are used as backscattering terms
in the edge theory, the former can be interpreted physically as a normal (charge-conserving)
backscattering between counterpropagating edge states, while the latter physically corresponds to superconductivity.
The resulting line algebra (\ref{magAlg}) in this special case becomes $W_{\b m_1}(a_i) W_{\b m_1'}(b_j) = W_{\b m_1'}(b_j) W_{\b m_1}(a_i) e^{\delta_{ij} 2\pi i 2/N}$.
When $N$ is odd, corresponding to fermionic FQH states, this gives a quantum dimension $d_{bulk} = \sqrt{N}$, which agrees with the
result $d = \sqrt{2} \sqrt{N}$ found in Ref.  \onlinecite{lindner2012,clarke2013,cheng2012}. As discussed in the previous subsection, the additional factor of $\sqrt{2}$ originates
from the purely 1D fermionic physics that gives rise to an additional Majorana fermion zero mode. When $N$ is even,
we have $d_{bulk} = \sqrt{N/2}$. The result $d = \sqrt{2N} = 2 \sqrt{N/2}$ of Ref. \onlinecite{lindner2012,clarke2013,cheng2012}, for $N$ even, has an additional
factor of $2$ releative to $d_{bulk}$. This is again due to purely one-dimensional bosonic physics, where there is an
additional boson parity symmetry of the model that was considered, which leads to additional ground state degeneracies. Breaking
this boson parity symmetry will lead to a topologically robust $\sqrt{N/2}$ quantum dimension.

As a second example, let us consider $Z_2$ twist defects of the state described by\cite{barkeshli2010,barkeshli2012a,barkeshli2013genon}
\begin{align}
K = \left( \begin{matrix} p & q \\ q & p \end{matrix} \right).
\end{align}
The special case $p = 0$ describes the case studied in Ref. \onlinecite{bombin2010,you2012}.
The twist defects have the property that a quasiparticle described by the vector $(q_1, q_2)$ is transformed into $(q_2, q_1)$ upon encircling the
twist defect. Such twists were shown to have a quantum dimension $\sqrt{|p-q|}$. Through the folding process, we obtain a
matrix
\begin{align}
\tilde{K} = \left(\begin{matrix} K & 0 \\ 0 & -K \end{matrix}\right).
\end{align}
The twist defect then maps to a domain wall
between gapped edges corresponding to two Lagrangian subgroups $M$ and $M'$. $M$ contains quasiparticles that are
generated by $\{\b m_1, \b m_2\} = \{(1,0,-1,0)^T, (0,1,0,-1)^T\}$ and $M'$ contains quasiparticles that are generated by
$\{\b m_1',\b m_2'\} = \{(1,0,0,-1)^T, (0,1,-1,0)^T\}$. In this basis, the algebra of Wilson line operators is described by (\ref{magAlg2}), with
\begin{align}
R = \frac{1}{p-q}\left(\begin{matrix} 1& -1 \\ -1 & 1 \end{matrix} \right).
\end{align}
Forming the lattice $\Gamma = \tilde{R} \mathbb{Z}^2$ (see eq. (\ref{RtildeDef})),
we obtain
\begin{align}
\tilde{R} = \frac{1}{p-q}\left(\begin{matrix} 1 & 0 \\ -1 & p-q \end{matrix} \right).
\end{align}
The quantum dimension of the defects is therefore $\sqrt{|Det \tilde{R}|^{-1}} = \sqrt{|p-q|}$, in agreement with previous
calculations.\cite{barkeshli2010,barkeshli2012a,barkeshli2013genon}

Finally, let us consider the boundary of $Z_N$ topological states, described by
\begin{align}
K = \left( \begin{matrix} 0 & N \\ N & 0 \end{matrix} \right).
\end{align}
As discussed in Sec. \ref{gappedEdgeEx}, two simple kinds of gapped edges correspond to condensation of either electric or magnetic particles.
These correspond to Lagrangian subgroups $M$ and $M'$ generated by $\b m_1 = (1,0)^T$ and $\b m_1' = (0,1)^T$, respectively.
The resulting Wilson line algebra is $W_{\b m_1}(a_i) W_{\b m_1'}(b_j) = W_{\b m_1'}(b_j) W_{\b m_1}(a_i) e^{\delta_{ij} i 2\pi /N}$. Therefore the domain wall
between these two kinds of gapped edges has a quantum dimension $\sqrt{N}$.

\subsection{Localized Toplogical Zero Modes}

A physical consequence of the existence of the domain walls is the presence of
a topologically robust non-zero density of states at zero energy
for a subgroup $L$ of the quasiparticles with fractional statistics. This subgroup $L$ is defined as follows
\begin{align}
\label{Ldef}
L = \{ \b m+\b m'| \b m \in M, \b m' \in M'\} .
\end{align}
The fractional statistics of the quasiparticles in $L$ derives from the fractional mutual statistics between
$\b m$ and $\b m'$. It will also be useful to define the subset of quasiparticles
\begin{align}
\tilde{L} &\equiv L \backslash (M \cup M')
\nonumber \\
&= \{ \b m+\b m'| \b m \in M, \b m' \in M', \b m+\b m' \notin M, M'\} .
\end{align}
$\tilde{L}$ simply contains the non-trivial quasiparticles in $L$ that do not belong to $M$ or $M'$.
We will show that the zero energy density of states of the quasiparticles in $\tilde{L}$ is exponentially localized
to the domain wall.

\begin{figure}
\centerline{
\includegraphics[width=2.3in]{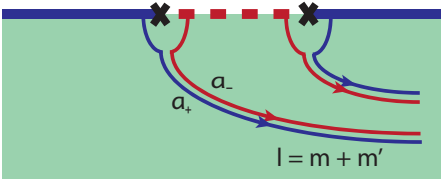}
}
\caption{
\label{zeroMode}
The domain walls between gapped edges labelled by Lagrangian subgroups $M$ and $M'$
allow for a process where an anyonic quasiparticle labelled $\b l = \b m+\b m'$, with $\b m \in M$, $\b m' \in M'$,
can be emitted from one domain wall and be absorbed at another. Schematically, the
$\b m$ quasiparticle is absorbed/emitted on one side of the domain wall,
while the $\b m'$ quasiparticle is absorbed/emitted on the other side. At the domain wall
both of these processes can occur together, allowing for the anyon $\b l = \b m + \b m'$ to be absorbed/emitted
at the domain wall. This directly implies that $\b l$ has a non-zero density of states at zero energy,
localized exponentially to the domain wall.
}
\end{figure}

To see this, first recall that the gapped regions introduce the processes
$W_{\b m}(a_i)$, $W_{\b m'}(b_i)$ (see Fig. \ref{lineAlg} (b)), which leave the system in the ground state subspace.
If we consider bringing the starting points of the paths near the domain walls, and pushing the end points out to infinity,
we have the process $W_{\b m + \b m'}(a) =\lim_{a_{\pm} \rightarrow a} W_{\b m}(a_+)W_{\b m'}(a_-)$ (see Fig. \ref{zeroMode}).
$a_\pm$ are the paths that start infinitesimally to the left/right of the domain wall and go out to infinity, and $a$ is the
limiting path obtained by fusing $a_+$ and $a_-$ together in such a way that they both start at the domain wall.
The quasiparticle $\b l$, which consists of the fusion of $\b m$ with $\b m'$,
can therefore be created at the position of the domain wall, propagate through the bulk,
and be annihilated at a different defect, keeping the system in the ground state subspace.
Away from the domain wall, $\b l$ cannot be absorbed or emitted, as long as $\b l \notin M,~ M'$
because it necessarily includes a non-trivial particle from both Lagrangian subgroups, and
only one Lagrangian subgroup is condensed on either side of the domain wall.
This directly implies that the anyons of the form $\b l = \b m + \b m'$ have a non-zero density of states, at zero energy,
localized to the domain walls, as long as $\b l \notin M,~M'$ (\it ie \rm $\b l \in \tilde{L}$).
These zero modes are topologically robust, as they are protected by the
topological nature of the gapped edges on either side of the domain wall.

\subsubsection{The zero mode in quasiparticle density of states}

\begin{figure}
\centerline{
\includegraphics[width=2.3in]{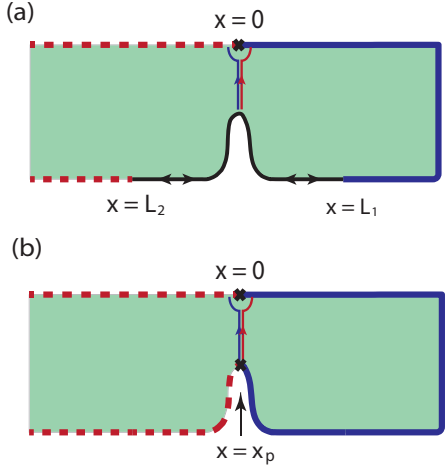}
}
\caption{Possible geometries for probing the quasiparticle zero energy density of states localized at the domain wall.
(a) There is a gapless edge in the region $x \in [L_1, L_2]$, from which a quasiparticle can tunnel into the zero mode
localized at the domain wall. (b) There can be a second domain wall at $x = x_p$, where the two defects at $x_p$ and $0$
can be arbitrarily far apart along the edge. A subset of the quasiparticles can be absorbed/emitted at zero energy
between domain walls that are well-separated along the edge.
\label{zeroMode2}
}
\end{figure}

In order to understand this more concretely in the edge theory let us consider a domain wall at $x = 0$
between $M$ and $M'$-edges. Furthermore, let us suppose that there is some region where the
edge is gapless, between $L_1 < x < L_2$, which we will use to ``probe'' the defect (Fig. \ref{zeroMode2}a).
Equivalently, the gapless region can be considered to shrink to a point (Fig. \ref{zeroMode2}b), in
which case we are considering the tunneling of quasiparticles from one defect to another.

More concretely, we consider the following backscattering terms on the edge:
\begin{align}
\delta H_{edge} =
g\sum_{i = 1}^N \left\{ \begin{array}{ccc}
\cos( c_i \b m_i^T \phi ) & \text{ for } & x\in [0,L_1] \\
\cos(c_i' \b m_i'^T \phi) & \text{ for }& x \in [x_L, 0], [L_2, x_R] \\
\end{array} \right.
\end{align}
where $x_L < 0$ and $x_R > L_2$. Recall that $\{\b m_i\}$ and $\{\b m_i'\}$ are the generators of $M$ and $M'$, respectively.
In the regions $x < x_L$ and $x > x_R$, we do not specify the nature of the edge, except to assume that it
allows topologically degenerate sectors due to the domain wall at $x = 0$. For concreteness
let us suppose all of the regions are infinitely long: $x_L \rightarrow -\infty$ and $L_1, L_2, x_R \rightarrow \infty$.
Furthermore, in the case where $L_1 < x < L_2$ is a gapless region of the edge theory, let us assume
$|L_1 - L_2| \rightarrow \infty$ in order for the edge to not have a finite-size gap in this region. Alternatively,
if $|L_1 - L_2| \rightarrow 0$, we can assume this point is the location of a second domain wall at
$x = L_1 = L_2 \equiv x_p$ (Fig. \ref{zeroMode2}b).

The classical minima in the $M$-gapped regions are set by $\b m_i^T \phi = 2\pi \alpha_i/c_i$, for integer $\alpha_i$.
The topologically degenerate sectors therefore can be (partially) labelled by
$|\vec{\alpha}\rangle$, such that  $e^{i \b m_i^T \phi(x)} |\vec{\alpha}\rangle = e^{i 2\pi \alpha_i/c_i} |\vec{\alpha}\rangle$
for $0 < x < L_1$. Note that in this basis, we cannot simultaneously diagonalize all of the
$e^{i\b m'^T \phi(x)}$ for $x_L < x < 0$, because of the non-trivial commutation relation between
$e^{i \b m^T \phi(y)}$ and $e^{i\b m'^T \phi(x)}$ at separate points $x$, $y$. Also note that this is a partial labelling of the states
because we have not specified the nature of the edge for $x < x_L$ and $x > x_R$.

Now, consider the quasiparticle operator
\begin{align}
\rchi_{\b l}(x) = e^{i \b l^T \phi(x)},
\end{align}
for $\b l \in \tilde{L}$. That is, for $\b l = \b m + \b m'$, with $\b m \in M$ and $\b m' \in M'$, such that $\b l \notin M, M'$.
We also consider $\b m$ and $\b m'$ to be generated by the null quasiparticle vectors $\{\b m_i\}$.
Consider the correlation function
\begin{align}
G_{\b l;\v \alpha}(x,t) = \langle \v \alpha |\rchi_{\b l}(x_p,t) \rchi_{\b l}^\dagger(x,t) \rchi_{\b l}(x,0) \rchi_{\b l}^\dagger(x_p,0)|\v \alpha\rangle,
\end{align}
where $|\v \alpha\rangle$ labels the different ground states under consideration. We include the operator $\rchi_{\b l}^\dagger(x_p,0)$, which
creates the quasiparticle $\b l$ at $x = x_p$, in order to be able to fully describe the system using only the Hilbert space in the long wavelength
theory of the edge, without reference to the bulk. $\rchi_{\b l}(x)$ alone is not a physical, gauge-invariant process on the edge. The point $x_p$ can be considered
to be located in the region of a gapless edge, or at another domain wall.

We can show that, at low frequencies, $\omega \ll g$, where $g$ is the scale of the energy gap of the edge states,
\begin{align}
\label{qpCorr}
G_{\b l;\v \alpha}(x,\omega) = \int dt G_{\b l;\v \alpha}(x,t) e^{i \omega t} \sim  e^{-|x|/\xi} \delta(\omega),
\end{align}
where $\xi \propto 1/g$ is the correlation length of the gapped edge states.
This directly implies that the quasiparticles of the form $\b l = \b m +\b m'$, such that $\b l \notin M,~M'$, have a non-zero density
of states, at zero energy, exponentially localized to the domain wall.
To see this, we insert a complete set of states:
\begin{align}
1 = \sum_a |a\rangle \langle a| = \sum_{\v \beta \in G} |\v \beta \rangle \langle \v \beta| + \sum_{n \in E} |n\rangle\langle n |,
\end{align}
where we have split the formal sum over all states in the edge theory into those in the ground state
subspace, labelled $G$, and the excited states, labelled $E$. Thus:
\begin{align}
G_{\b l;\v \alpha}(x,t) = \sum_{\v \beta\in G} |\langle \v \beta | \rchi_{\b l}(x) \rchi_{\b l}^\dagger(x_p) |\v \alpha\rangle|^2 + \cdots
\end{align}
The $\cdots$ represent the sum over excited states, which can be neglected at frequencies much smaller than
the energy gap in the edge states. Now consider the matrix elements
\begin{align}
\label{qpMatrixEl}
|\langle \v \alpha | \rchi_{\b l}(x) \rchi_{\b l}^\dagger(x_p) |\v \beta \rangle| = |\langle \v \alpha | e^{i \b m'^T (\phi(x) - \phi(x_p))}| \v \beta \rangle|,
\end{align}
where we have assumed, without loss of generality, that $x > 0$, so that $e^{i \b m^T \phi(x)}$ are diagonalized
in the $\v \alpha$ basis, and thus simply contribute an unimportant $U(1)$ phase factor to the matrix elements.

Furthermore, observe that the operators $e^{i \b m'^T (\phi(0) - \phi(x_p))}$, for $\b m' \in M'$ act as ladder operators
that connect the different ground states $|\v \alpha\rangle$. To see this explicitly
consider the commutation relations for $x > 0$:
\begin{align}
[\b m^T \phi(x), \b m'^T(\phi(0) - \phi(x_p))] = i 2\pi \b m^T K^{-1} \b m',
\end{align}
which implies that
\begin{align}
e^{i \b m'^T (\phi(0) - \phi(x_p))}|\v \alpha \rangle = |\v \gamma\rangle,
\end{align}
where $\gamma_i = \alpha_i + \b m'^T (c_i K^{-1} \b m_i)$. Therefore we can write:
\begin{align}
|\v \alpha\rangle = e^{-i \b m'^T (\phi(0)-\phi(x_p)) } |\v \gamma \rangle.
\end{align}
Eq. (\ref{qpMatrixEl}) thus simplifies to
\begin{align}
|\langle \v \alpha | \rchi_{\b l}(x) \rchi_{\b l}^\dagger(x_p) |\v \beta \rangle| = |\langle
\v \gamma| e^{-i \b m'^T \phi(0)  } e^{i \b m'^T \phi(x)  }  |\v \beta \rangle|.
\end{align}
The operators inside the expectation value simply introduce a kink in the uniform boson
configuration at position $x$. It is clear that this equal-time correlation function
must decay exponentially:
\begin{align}
\label{matrixEl2}
|\langle \v \gamma| e^{-i \b m'^T \phi(0)} e^{i \b m'^T \phi(x)}  |\v \beta \rangle| \propto \delta_{\v \gamma \v \beta} e^{-x/\xi} , \;\; x > 0
\end{align}
because the edge states are gapped for $x > 0$ and $x < 0$, with a correlation length $\xi \propto 1/g$.
The Kronecker delta $\delta_{\v \gamma \v \beta}$ arises because the $M$-edge extends to infinity, while the kink
is created at a finite distance $x$ away from the location of the domain wall.

We can repeat a slightly modified argument to obtain an exponential decay for $x < 0$ as well.
This proves (\ref{qpCorr}), and therefore that the quasiparticles in $\tilde{L}$ have a non-zero density of states at
zero energy, exponentially localized to the domain wall.

We see that the physical origin of this non-zero density of states is simply that
$\rchi_{\b l}(0) \rchi_{\b l}^\dagger(x_p)$ can take one degenerate ground state to another: the domain wall
can absorb/emit quasiparticles of the form $\b l = \b m + \b m'$, while staying in the ground state subspace.
The $\delta(\omega)$ arises because, aside from the degenerate ground state subspace, there is a gap
of order $g$ to any other states, and therefore for $\omega \ll g$ there will be no spectral weight
aside from the delta function at $\omega = 0$.

While the $\rchi_{\b l}$ for $\b l \in \tilde{L} $ have a non-zero density of states at zero energy localized
to the domain wall, the zero-energy density of states is not localized 
if $\b l \in M$ or $M'$. In contrast, if $\b l \notin L$, then there will be a vanishing density of states at zero energy everywhere, because
such quasiparticle operators do not act in the degenerate ground state subspace, even when applied at
the domain wall.

In the fermionic case, where $K$ has at least one odd element along the diagonal, there is an additional subtlety. As stated in
Sec. \ref{proofSec}, we can always pick a choice of $K$ and a basis $\{\b m_i\}$ for the Lagrangian subgroup $M$,
such that $\b m_i^T K^{-1} \b m_j = 0$. While $\b m_i$ and $\b m_i + K\b \Lambda$, for $\b \Lambda$ an integer vector, describe the same topological
quasiparticle, they may differ in fermion parity: One may be a boson while the other is a fermion.
In this case, $\rchi_{\b l}(x)$ has a robust zero energy density of states only if $\b l = \b m+\b m'$, with $\b m$, $\b m'$ both bosons.
Otherwise, creating the quasiparticle $\b l$ on the edge will require also adding an additional local
fermion, which will not be guaranteed to have a non-zero density of states at zero energy, unless there happens to be an
additional Majorana fermion zero mode due to purely one-dimensional physics, as discussed in Sec. \ref{qdimDiscSec}. Indeed,
the above analysis assumed explicitly that one can choose a basis where either $e^{i \b m^T \phi(x)}$ or $e^{i \b m'^T \phi(x)}$  can
acquire non-zero expectation values for $x > 0$ and $x < 0$, respectively, which is only possible for bosonic operators.

\subsubsection{Generalized parafermion zero mode algebra}

Now let us consider a set of domain walls between the $M$ and $M'$-gapped regions, at
the positions $x_i$. Based on the above analysis, the quasiparticle operators $\rchi_{\b l}(x)$ have a
non-zero local density of states at zero energy at $x_i$, for quasiparticles $\b l \in L$.
We define the zero mode operators (see. Fig. \ref{zeroMode}):
\begin{align}
\label{pSplit}
\rgamma_{\b li} = \lim_{\epsilon \rightarrow 0^+} e^{i \b m^T \phi(x_i + \epsilon)} e^{i \b m'^T\phi(x_i - \epsilon)},
\end{align}
for $\b l  =\b m + \b m'$, if $x_i + \epsilon$ is an $M$-gapped region and $x_i - \epsilon$ is an $M'$-gapped region.
For the reverse scenario, the role of $\b m$ and $\b m'$ are interchanged above.
This regularization is important in order to properly define the commutation relations between different $\rgamma_{\b li}$ on the same domain wall.

Note that the operators $\rgamma_{\b li}$ are zero mode operators in the sense that they can be created/absorbed at the domain walls
at zero energy, as long as they are correspondingly absorbed/created somewhere else.
That is, $\rgamma_{\b li}^\dagger \rgamma_{\b lj}$ always keep the edge in its ground state subspace and therefore commute with the Hamiltonian:
\begin{align}
[H_{edge}, \rgamma_{\b li}^\dagger \rgamma_{\b lj} ] = 0.
\end{align}
In general, the $\rgamma_{\b li}$ individually either commute or anti-commute with $H_{edge}$, since they are mutually local.

The zero modes $\rgamma_{\b l;i}$ satisfy the following algebra:
\begin{align}
\label{pfAlg1}
\rgamma_{\b l;i} \rgamma_{\b{\tilde{l}};j} &= \rgamma_{\b{\tilde{l}};j} \rgamma_{\b l;i} e^{i \pi \b l^T K^{-1} \b{\tilde{l}} sgn(i - j) } , \;\; i \neq j,
\end{align}
\begin{align}
\label{pfAlg2}
\rgamma_{\b l;i} \rgamma_{\b{\tilde{l}};i} &= \rgamma_{\b l + \b{\tilde{l}};i} e^{\pm i \pi \b m' K^{-1} \b{\tilde{m}}}
\nonumber \\
&= \rgamma_{\b{\tilde{l}};i} \rgamma_{\b l;i} e^{\pm i \pi (\b m'^T K^{-1} \b{\tilde{m}} - \b m K^{-1} \b{\tilde{m}')}},
\end{align}
where $\b l = \b m + \b m'$ and $\b{\tilde{l}} = \b{\tilde{m}} + \b{\tilde{m}}'$, with $\b m,\b{\tilde{m}} \in M$ and $\b m', \b{\tilde{m}}'\in M'$.
The $\pm$ sign in the latter equation depends on whether the domain wall has an $M$-edge to the left and an $M'$-edge
to the right, or vice versa, and can be obtained using the point-splitting regularization defined in (\ref{pSplit}).
Each operator $\rgamma_{\b l;i}$ has a finite order: $\rgamma_{n_l \b l}$ will commute with the whole algebra for some integer $n_l$,
and therefore can be represented in the algebra by the identity operator.

Eq. (\ref{pfAlg1}), (\ref{pfAlg2}) define a \it generalized parafermion \rm algebra. The simplest version of this
algebra, which consists of a single parafermion generator,\cite{fradkin1980,zamolodchikov1985,fendley2012} appear
for the zero modes localized to the topological defects considered in
Ref. \onlinecite{barkeshli2012a,barkeshli2013genon,you2012,lindner2012,clarke2013,cheng2012,vaezi2013,barkeshli2013,you2013}.

\subsection{Mapping to genons}
\label{genonDisc}

The Wilson line algebra (\ref{magAlg}) induced by the defects is similar to the Wilson loop algebra
of an Abelian CS theory on a high genus surface. In the latter case, the Wilson loop algebra describes
the quasiparticle propagation along non-contractible cycles of the high genus surface. Here, we will
show that (\ref{magAlg}) is exactly equivalent to the Wilson loop algebra of some Abelian
CS theory on a high genus surface. Consequently, we refer to such boundary defects as \it genons \rm,
in the sense of topological equivalence.

Let us begin by considering the following algebra for the Wilson lines:
\begin{align}
\label{magAlg3}
A_{\vec q} B_{\vec{q'}} = B_{\vec{q'}} A_{\vec q}  e^{i 2\pi \vec{q}^T \tilde{R} \vec{q'}} ,
\end{align}
where recall $\vec{q}$ and $\vec{q'}$ are $N$-component integer vectors. From the definition
of $\tilde{R}$ (see eq. (\ref{RtildeDef}) and subsequent discussion), we can see that this algebra is
equivalent to the original algebra (\ref{magAlg2}) up to a possible relabelling of the operators.

Here we would like to show that one can always choose a basis of the Lagrangian subgroups
$M$ and $M'$ so that $\tilde{R}^{-1}$ is a diagonal integer matrix.
To see that it is integer, observe that
the lattice $\Gamma$ includes every integer vector $\v \Lambda \in \mathbb{Z}^{N}$. Therefore,
$\forall \v \Lambda \in \mathbb{Z}^{N}$, there must exist $\v \Lambda' \in \mathbb{Z}^{N}$ such
that $\tilde{R} \v \Lambda' = \v \Lambda$. In other words, for every $\v \Lambda \in \mathbb{Z}^{N}$,
$\tilde{R}^{-1} \v \Lambda$ is an integer vector, which implies that $\tilde{R}^{-1}$ must be integer.

To see that $\tilde{R}^{-1}$ can be diagonal, we use the following theorem of linear algebra. If $A$ is an
integer $N \times N$ matrix, then there exist integer matrices $S$ and $T$ with unit determinant
such that $A' = S A T$ is a diagonal integer matrix. $A'$ is known as the \it Smith normal form \rm
of $A$.\cite{prasolov1994}
Applying this to $\tilde{R}$ implies that one can always find a basis of Lagrangian subgroups for $M$ and $M'$ such that
$\tilde{R}$ will be diagonal.

Therefore if we define $\tilde{K} \equiv \tilde{R}^{-1}$, then (\ref{magAlg3}) can be interpreted as the
Wilson loop algebra of a $U(1)^{N}$ CS theory on a torus, characterized by the matrix $\tilde{K}$, where
$A_{\vec q}$ and $B_{\vec q}$ are represented as
\begin{align}
A_{\vec q} = e^{i q_I \oint_a a_I \cdot dl}, \;\; B_{\vec q} = e^{i q_I \oint_b a_I \cdot dl}.
\end{align}
Here, $a_I$ are the $U(1)$ gauge fields of this $U(1)^{N}$ CS theory,
$a$ and $b$ are the non-contractible loops on the torus, and $a + b$ is the non-contractible
loop that encircles both $a$ and $b$ once.

As is well known, such a theory has a topological degeneracy given by $|\text{Det } \tilde{K}| = |\text{Det } \tilde{R}|^{-1} = D$.
Therefore with $2n$ domain walls, there are $n-1$ copies of the above algebra, and therefore the algebra corresponds to
the Wilson loop algebra of the associated $U(1)^{N}$ CS theory with matrix $\tilde{K}$ on a genus $g = n-1$ surface.

We will show in Sec. \ref{braidSec} that one can also define a notion of braiding of these domain walls,
which correspond to modular transformations, or Dehn twists, of the corresponding Abelian CS theory on the genus
$g = n-1$ surface. This dramatically generalizes earlier results
about twist defects\cite{barkeshli2013genon}. Since the ground state degeneracy and braiding of the defects in these
cases can be understood in terms of the properties of an Abelian CS theory on a high genus surface, we refer to
these defects as \it genons\rm.

We would also like to note that
the defect and anti-defect are topologically equivalent. In the basis in which $\tilde{R}$ is symmetric, the Lagrangian subgroups $M$ and $M'$ can be mapped to
each other, with a quasiparticle ${\b m}=q_i{\b m}_i\in M$ being mapped to ${\b m}'=q_i{\b m}_i'\in M'$. Using this mapping,
the on-site commutation relations of the parafermion zero modes (\ref{pfAlg2}) are the same for defects and anti-defects.
This is consistent with the case of $Z_2$ twist defects studied previously,\cite{barkeshli2013genon} where defects and
anti-defects are equivalent to each other.

\subsection{Projective Non-Abelian Statistics}
\label{braidSec}

We have seen that the domain walls of the gapped edge states localize topologically
protected zero modes, and give rise to topological degeneracies. This raises the
question of whether it is possible to ``braid'' these defects; that is, to carry out topologically
protected unitary transformations in the degenerate subspace.

Since the defects generally exist as domain walls on the boundary of a topological phase, it is not
possible to geometrically braid the defects and return to the original configuration of the system.
However, it has been shown that in some cases it is possible to define a notion of
braiding, through a different approach that involves tuning the tunneling of quasiparticles
between the domain walls. \cite{alicea2010b,lindner2012,clarke2013,barkeshli2013genon,bonderson2013}

In particular, let us consider bringing the edges close together, in order to induce quasiparticle tunneling between either
the domain walls or the gapped domains. This leads to a Hamiltonian that acts on the degenerate ground
state subspace, opening the possibility of finding a closed path in Hamiltonian space that
successfully carries out an adiabatic non-abelian Berry phase on the ground state
subspace. Such non-Abelian Berry phases can, under certain conditions, be topologically protected
up to an overall phase. This yields the possibility of defining a notion of projective non-Abelian statistics
for the defects.

In what follows, we will show 
it is in general possible to generate the following topologically
protected transformations. Consider an array of point defects, labelled $1, \cdots, 2n$, separating
$M$ and $M'$-edges. Pick any neighboring pair of defects, such as
$1$ and $2$ for convenience, and define the loops $\{a_i\}$ and $\{b_i\}$, as shown in Fig. \ref{lineAlg} (b).
Without loss of generality, we suppose that defect $1$ has an $M$-gapped edge to the left and an $M'$-gapped edge to the right,
and vice versa for defect $2$.

We will show that we can generate a topologically protected ``braiding'' transformation $B_{12}$,
which is a unitary transformation in the Hilbert space of topologically degenerate states, and has the
following  action on the Wilson line operators $\{W_{\b m}(a_i)\}$ and $\{W_{\b m'}(b_i)\}$:
\begin{align}
\label{braiding}
B_{12}: & W_{\b m}(a_{1}) \rightarrow e^{i\theta_{\b m}} W_{\b m}(a_{1})
\nonumber \\
& W_{\b m'}(b_{1}) \rightarrow e^{i\theta'_{\b m}} W_{\b m}^\dagger(a_1) W_{\b m'}(b_1),
\end{align}
while the rest of the Wilson line operators are left invariant. The phases
$e^{i\theta_{\b m}}$ and $e^{i\theta'_{\b m}}$ will be defined below, in eq. (\ref{braidingPhaseDef}).
Eq. (\ref{braiding}) requires a canonical pairing between quasiparticles $\b m\in M$ and $\b m' \in M'$,
which we will explain below. In the mapping to the high genus surface (see Sec. \ref{genonDisc}), the
above braiding transformations can simply be understood as Dehn twists, or modular transformations,
of the genus $g$ surface. This generalizes the result found in Ref. \onlinecite{barkeshli2013genon} in the
context of twist defects to a more general class of point defects of an Abelian topological phase.

In order to derive the above result, we first define the following zero mode Hamiltonian:
\begin{align}
\label{HabDef}
H_{ab} = \sum_{i=1}^N (t_{\b l_i} \rgamma_{\b l_i;a}^\dagger \rgamma_{\b l_i;b} +H.c.)
\end{align}
Here, the quasiparticle $\b l_i$ is defined by
\begin{align}
\label{liDef}
\b l_i = \b m_i + \b m_i', \;\; i = 1, \cdots, N,
\end{align}
where $\{\b m_i\}$ and $\{\b m_i'\}$ for $i = 1,.., N$ are the generators of $M$ and $M'$, in the basis defined by
eq. (\ref{magAlg3}) with $\tilde{R}$ diagonal. Note that this basis pairs every generator $\b m_i$ of $M$ with a generator
$\b m_i'$ of $M'$, and therefore induces a natural pairing between every $\b m \in M$ and $\b m' \in M'$.
In the notation of (\ref{magAlg3}), this is the statement that for every $N$-dimensional vector
$\v q$, we define an operator $A_{\v q} \equiv W_{\sum_i q_i \b m_i}(a_1)$, which is associated to Wilson lines of particles in $M$,
and an operator $B_{\v q} \equiv W_{\sum_i q_i \b m_i'}(b_1)$, which is associated to Wilson lines of particles in $M'$.
Recall that $N$ is half the dimension of the $K$-matrix: $dim(K) = 2N$,
and, following the discussion of Sec. \ref{proofSec}, we are picking $N$ generators
$\{\b m_i\}$ of $M$, and similarly for $M'$.
Note that the individual terms in the sum, $\rgamma_{\b l_i;a}^\dagger \rgamma_{\b l_i;b} $,
commute with each other.

\begin{figure}
\centerline{
\includegraphics[width=2.5in]{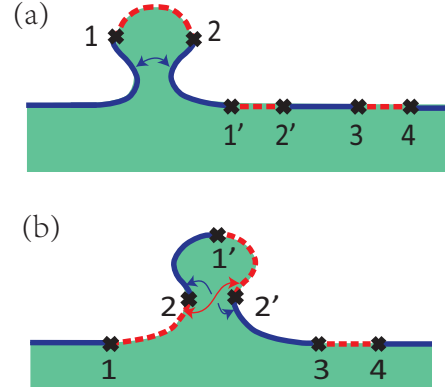}
}
\caption{
\label{HabFig}
(a) Bringing different gapped edges together can induce tunneling of quasiparticles associated with the corresponding
Lagrangian subgroups. $H_{12}$ can for example be realized using the geometry shown, with the double arrows indicated
tunneling of an $\b m \in M$ quasiparticle between $M$-gapped edges (blue solid lines).
(b) Bringing different defects (such as $2$ and $2'$) in close proximity can induce tunneling of quasiparticles with zero modes localized to the defects.
Double arrows indicate tunneling of zero modes between defects.
}
\end{figure}

When $a$ and $b$ are nearest neighbors, \it eg \rm $b = a+1$, the sum in $H_{a,a+1}$
effectively runs over all possible zero modes. This is equivalent to a sum over all Wilson line
operators that connect the gapped region to the left of the defect at $x_a$ and the region to the right of the
defect at $x_{a+1}$. Physically, this can be achieved by bringing the edges in close physical proximity, as in
Fig. \ref{HabFig}a.

In contrast, when $a$ and $b$ are not nearest neighbors, then we see that $H_{ab}$ only sums over a restricted
set of zero modes. For the braiding we define below to be topologically robust, it is crucial that
the only terms which appear in the sum consist of quasiparticles generated by those of the form
$\b{l}_i = \b{m}_i + \b{m}_i'$. Couplings of any other zero modes must be exponentially suppressed.
In general, zero modes at $x_a$ and $x_b$ can be coupled by bringing the point defects at $x_a$ and $x_b$
in close proximity in order to induce the relevant quasiparticle tunneling (see Fig. \ref{HabFig}b).
However, in order to suppress the tunneling of quasiparticles that are not generated by
$\{\b{l}_i\}$, there may need to be additional geometric or energetic constraints. An example of
such a geometric constraint occurs in the examples studied in Ref. \onlinecite{lindner2012,clarke2013},
where physically the system is not folded (in the sense of Sec. \ref{foldingSec}), and bringing well-separated, non-neighboring
defects together will only allow a subset of the zero modes to tunnel with appreciable amplitude from one
defect to another through the bulk.

\begin{figure}
\centerline{
\includegraphics[width=2.9in]{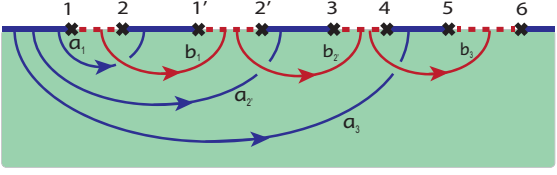}
}
\caption{
\label{braid1}
For the braiding between defects $1$ and $2$, we introduce another ancillary pair of defects $1'$ and $2'$.
The degrees of freedom at $1$ can then first be transferred to $2'$, and then those at $2$ can be transferred to
$1'$. After the process, $1$ and $2$ are coupled and can be annihilated, and $1'$ and $2'$ will be relabelled
$1$ and $2$, respectively. This effectively exchanges the degrees of freedom at $1$ and $2$.
}
\end{figure}
In order to ``braid'' the defects $1$ and $2$, we introduce another pair of domain walls, $1'$ and $2'$ (see Fig. \ref{braid1}),
which are initially coupled via $H_{1'2'}$ and therefore do not contribute to the initial ground state degeneracy.
Subsequently, we consider the following path in Hamiltonian space:
\begin{align}
H(\tau) = \left\{ \begin{array}{cc}
H_{2\rightarrow 1'} = (1-\tau) H_{1'2'} + \tau H_{22'},  & \tau \in [0,1] \\
H_{2\rightarrow 1'} = (2 - \tau) H_{22'} + (\tau - 1) H_{12}, & \tau \in [1,2]\\
\end{array} \right.
\end{align}
Focusing on the $D^2$-dimensional Hilbert space associated with the defects $1$, $2$, $1'$, and $2'$
(see Sec. \ref{qdimSec} for a definition of $D$), we can show that $H_{1'2'}$, $H_{22'}$, and $H_{12}$ will
have a ground state degeneracy of $D$, with a gap to the next excited states, assuming certain
special values of the hopping coefficients $t_{\b{l}_i}$ are avoided (see Appendix \ref{gsdAppendix}
for a detailed explanation). Therefore, we expect that $H(\tau)$ will have $D$ ground states throughout the process;
if there are any additional accidental degeneracies along the way, it is possible to choose a slightly different path in Hamiltonian space
that avoids them without modifying $H(0)$, $H(1)$, and $H(2)$. Therefore the system is in the $D$
dimensional ground state subspace throughout the entire adiabatic process.

The first process $H_{2\rightarrow 1'}$ can be thought of as transferring the zero modes from $2$ to $1'$,
while the second process $H_{1 \rightarrow 2'}$ can be thought of as transferring the zero modes from $1$ to
$2'$. After the process is over, we again have two uncoupled domain walls, $1'$ and $2'$. Therefore
$1'$, $2'$ after the process play the role of $1$, $2$ before the process. Therefore if we rename $1'$
and $2'$ to be the new $1$ and $2$, we can see that $1$ and $2$ are effectively exchanged.

To be more explicit, we define the following physical operators\cite{clarke2013,lindner2012,barkeshli2013genon}:
\begin{align}
\mathcal{O}_{\b l;1} = \rgamma_{\b l;2} \rgamma^\dagger_{\b l;1'} \rgamma_{\b l;2'} \rgamma_{\b l;\infty}^\dagger
\nonumber \\
\mathcal{O}_{\b l;2} = \rgamma_{\b l;1}^\dagger \rgamma_{\b l;2} \rgamma_{\b l;2'}^\dagger \rgamma_{\b l;\infty}.
\end{align}
Here we have included the operator $\rgamma_{\b l;\infty}$, which represents a quasiparticle operator
at some reference defect (or, alternatively, infinitely far away), which is necessary to ensure
that $\mathcal{O}_{\b l;1}$ and $\mathcal{O}_{\b l;2}$ are physical operators that act on the Hilbert
space of the edge theory.

The action of $H(\tau)$ on the ground state subspace can be understood by noticing that
these operators commute with the two processes:
\begin{align}
[H_{2 \rightarrow 1'}, \mathcal{O}_{\b l_i;1}] &= 0
\nonumber \\
[H_{1 \rightarrow 2'}, \mathcal{O}_{\b l_i;2}] & = 0,
\end{align}
for $\b l_i$, $i = 1,\cdots, N$, of the form described in (\ref{liDef}).
The ground states of $H_{ab}$ have definite eigenvalues for the operators
$\rgamma_{\b l_i;a}^\dagger \rgamma_{\b l_i;b}$. For $(a,b) =  (1,2), (1',2'),(2,2')$, we have:
\begin{align}
\rgamma_{\b l_i;a}^\dagger \rgamma_{\b l_i;b} |k;\v  \alpha_{ab}\rangle = e^{i 2\pi (\v \alpha_{ab})_i} |k; \v \alpha_{ab}\rangle.
\end{align}
$k = 1,\cdots, D$ parametrize the $D$ ground states of $H_{ab}$,
while $\v \alpha_{ab}$ characterize the eigenvalues of $\rgamma_{\b l_i;a}^\dagger \rgamma_{\b l_i;b}$.

Defining $\mathcal{P}(\tau)$ to be the projection onto the ground state subspace of $H(\tau)$, we find:
\begin{align}
\mathcal{P}(0) \mathcal{O}_{\b l_i;1} \mathcal{P}(0) &= e^{i 2\pi (\v \alpha_{1'2'})_i} \rgamma_{\b l_i;2} \rgamma_{\b l_i;\infty}^\dagger
\nonumber \\
\mathcal{P}(1) \mathcal{O}_{\b l_i;1} \mathcal{P}(1) &= e^{-i2\pi (\v \alpha_{22'})_i} e^{i\pi{\b l}_i^T K^{-1} {\b l}_i} \rgamma_{\b l_i;2'}  \rgamma^\dagger_{\b l_i;1'} \rgamma_{\b l_i;2'} \rgamma_{\b l_i;\infty}^\dagger
\end{align}
This implies that after the first process, $\tau =  0 \rightarrow 1$, the zero mode $\rgamma_{\b l_i;2}$
transforms into $e^{-2\pi i ((\v \alpha_{1'2'})_i + (\v \alpha_{22'})_i)} e^{i\pi{\b l}_i^T K^{-1} {\b l}_i} \rgamma_{\b l_i;2'} \rgamma^\dagger_{\b l_i;1'} \rgamma_{\b l_i;2'}$.

Similarly,
\begin{align}
\mathcal{P}(1) \mathcal{O}_{\b l_i;2} \mathcal{P}(1) &= e^{-i2\pi (\v \alpha_{22'})_i} e^{i\pi{\b l}_i^T K^{-1} {\b l}_i} \rgamma_{\b l_i;1}^\dagger \rgamma_{\b l_i;\infty},
\nonumber \\
\mathcal{P}(2) \mathcal{O}_{\b l_i;2} \mathcal{P}(2) &= e^{i 2\pi (\v \alpha_{12})_i}\rgamma_{\b l_i;2'}^\dagger \rgamma_{\b l_i;\infty},
\end{align}
which implies that after the second process, $\tau = 1 \rightarrow 2$,
$\rgamma_{\b l_i;1}^\dagger \rightarrow e^{-i\pi{\b l}_i^T K^{-1} {\b l}_i} e^{i 2\pi ( ( \v \alpha_{12})_i + (\v \alpha_{22'})_i)} \rgamma_{\b l_i;2'}^\dagger$.
Also, we note that
\begin{align}
[\rgamma_{\b l_i;2'} \rgamma_{\b l_i;1'}^\dagger \rgamma_{\b l_i;2'} \rgamma_{\b l_i;\infty}^\dagger, H_{1 \rightarrow 2'}] = 0,
\end{align}
so that the result of the first process is not affected by the second process.

At the beginning of this process, we had the zero modes at $x_1$ and $x_2$ while
the ones at $x_1'$ and $x_2'$ were coupled and therefore did contribute to the ground state degeneracy.
At the end of the process, we ended up with the zero modes at $x_1'$ and $x_2'$ while the ones at
$x_1$ and $x_2$ are coupled. The final system is equivalent to the original system, and
so we can just relabel $\rgamma_{\b l_i;1'}$ and $\rgamma_{\b l_i;2'}$ as $\rgamma_{\b l_i;1}$ and $\rgamma_{\b l_i;2}$, respectively.

The above results imply that the braiding $B_{12}$ has the following action on the zero mode operators:
\begin{align}
B_{12}^\dagger \rgamma_{\b l_i;2} B_{12} &= e^{-2\pi i ( (\v \alpha_{1'2'})_i + (\v \alpha_{22'})_i)} e^{i\pi{\b l}_i^T K^{-1} {\b l}_i}\rgamma_{\b l_i;2} \rgamma_{\b l_i;1}^\dagger \rgamma_{\b l_i;2}
\nonumber \\
B_{12}^\dagger \rgamma_{\b l_i;1} B_{12} &= e^{-i 2\pi ( (\v \alpha_{12})_i + (\v \alpha_{22'})_i)}  e^{i\pi{\b l}_i^T K^{-1} {\b l}_i}\rgamma_{\b l_i;2}
\end{align}

In terms of the Wilson lines,
\begin{align}
W_{\b m_i}(a_1) &= \rgamma_{\b l_i;1}^\dagger \rgamma_{\b l_i;2} ,
\nonumber \\
W_{\b m_i'}(b_1) &= \rgamma_{\b l_i;2}^\dagger \rgamma_{\b l_i;3},
\end{align}
we obtain the result (\ref{braiding}) for the action of $B_{12}$, with the phases
\begin{align}
\label{braidingPhaseDef}
e^{i\theta_{\b m} } = e^{2\pi i \v q \cdot (\v \alpha_{12} - \v \alpha_{1'2'} )},
\nonumber \\
e^{i\theta'_{\b m}} = e^{2\pi i \v q \cdot (\v \alpha_{1'2'} + \v \alpha_{22'})},
\end{align}
where the $N$-component integer vector $\v q$ is defined by
$\b m = \sum_{i=1}^N q_i \b{m}_i$.

Using these results, we can explicitly derive the braid matrix, in a given basis.
As explained in Sec. \ref{qdimSec}, we can label the ground states associated with the
defects $1$, $2$ in terms of eigenvalues $\v \alpha$ of $A_{\vec q} \equiv W_{\sum_i q_i m_i} (a_1)$:
$A_{\vec q} |\v \alpha \rangle = e^{2\pi i \v q \cdot \v \alpha} |\v \alpha \rangle$.
In this basis, $B_{\vec q} \equiv W_{\sum_i q_i m_i'}(b_1)$ act as ladder operators, so that
\begin{align}
|\v \alpha \rangle = B_{\vec q} |0 \rangle,
\end{align}
where
\begin{align}
\label{alphaRel}
\v \alpha = \tilde{R} \v q
\end{align}
(see eq. (\ref{RtildeDef}) for a definition of $\tilde{R}$). Therefore, the braid matrix is:
\begin{align}
B_{12}^\dagger |\v \alpha \rangle &= B_{12}^\dagger B_{\vec q} |0\rangle = e^{i \phi} B_{12}^\dagger B_{\v q} B_{12} |0\rangle
\nonumber \\
&= e^{i\phi + i \theta'_{\b m}} A_{\v q}^\dagger B_{\v q} |0 \rangle = e^{i\phi + \theta'_{\b m}} e^{-i2\pi \v q^T \tilde{R} \v q} |\alpha \rangle.
\end{align}
Here, $e^{i\phi}$ is an undetermined phase, associated with the eigenvalue $B_{12}^\dagger |0 \rangle$,
and $\b m$ is defined by $\v q$: $\b m = \sum_i q_i \b m_i$.
Therefore, in this basis,
\begin{align}
(B_{12}^\dagger)_{\v q \v{q'}} = \delta_{\v q \v{q'}} e^{i\phi} e^{i 2\pi i \v q \cdot (\v \alpha_{1'2'} + \v \alpha_{22'}) } e^{-i2\pi \v q^T \tilde{R} \v q}.
\end{align}
$\v q$ and $\v{q'}$ are related to the eigenvalues $\v \alpha$ and $\v \alpha'$ via $\tilde R$, as shown in
(\ref{alphaRel}).

The above sequence of Hamiltonians can be thought of as inducing interactions between the defects, which
leads to the quasiparticle tunneling between the defects. As discussed in Ref. \onlinecite{bonderson2013},
adiabatically tuning such interactions is equivalent to performing a sequence of pairwise projections on the
subspace associated with different pairs of defects. Various sequences of such projections
can lead to a non-trivial unitary transformation on the original space of ground states and effectively carries out
a braiding process. Since the time-dependent Hamiltonian simply realizes these projection operators,
it is clear that small deformations of the path in Hamiltonian space will not affect the result except up
to an overall phase, as long as $H(\tau)$, when considered in the Hilbert space of the defects $1$, $2$, $1'$, $2'$,
has a $D$-dimensional ground state degeneracy throughout entire the process, with a finite gap to other excited states.
Therefore  the resulting non-abelian Berry phase is \it topologically protected. \rm
Since the overall phase is not topological, we refer to this as a \it projective \rm realization of
non-Abelian statistics. That the braiding operations indeed satisfy projectively the defining relations of the braid group
follows from the fact that we have shown they can be mapped to modular transformations of an Abelian CS
theory on a high genus surface, which are known to form a projective representation of the braid group.

\section{Critical phenomena between gapped edge states}
\label{transSec}

A defining property of topologically distinct gapped edges is that it is not possible to adiabatically tune from
one to the other without closing the energy gap on the edge. This raises the question of whether we can
understand the critical phenomena between different gapped edges in terms of the topological properties
of the different gapped edges.

\begin{figure}
\centerline{
\includegraphics[width=3.0in]{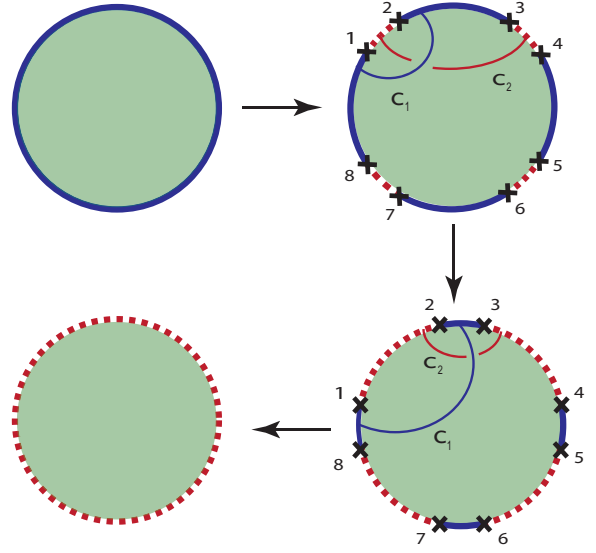}
}
\caption{\label{transFig} Starting with a gapped edge corresponding to a Lagrangian subgroup $M$, we can nucleate
pairs of domain walls enclosing gapped regions corresponding to a Lagrangian subgroup $M'$,
separate them, and re-annihilate them in pairs, leaving the $M'$ gapped edge. In the $M$-phase, the dominant
quasiparticle tunnelings are along the $c_{2i-1}$ paths enclosing the $M'$ regions. In the $M'$ phase,
the dominant quasiparticle tunnelings are along the $c_{2i}$ paths enclosing the $M$ regions.
}
\end{figure}

Let us focus on the transition between two different kinds of gapped edges, labelled by Lagrangian subgroups
$M$ and $M'$. One way to understand a transition between the $M$ edge and the $M'$ edge is as follows.
Consider starting with the $M$-edge and nucleating $N$ pairs of point defects that enclose the $M'$-edge,
as shown in Fig. \ref{transFig}. Next, the size of the $M'$ regions is increased, until the $M$ regions shrink to
zero. This process can be described by the following Hamiltonian, which acts in the topologically degenerate
subspace of the $N$ pairs of point defects:
\begin{align}
\label{spinChainH}
H_{edge} &= \sum_{i=1 }^N (A_i + B_i)
\nonumber \\
A_i &= \sum_{\b m\in M} t_{\b m} W_{\b m}(c_{2i-1})  + H.c.
\nonumber \\
B_i &= \sum_{\b m'\in M'} \tilde{t}_{\b m'} W_{\b m'}(c_{2i}) + H.c.,
\end{align}
where $i = 2N+1$ and $i=1$ label the same domain wall. The paths $c_i$ are defined to enclose
the defects $i$ and $i+1$, as shown in Fig. \ref{transFig}.

Here we have included only Wilson lines that enclose one pair of defects to obtain a model Hamiltonian
that can describe the transitions between different edge states. A Hamiltonian that is more physically
realistic for a given microscopic setup may also include longer-range tunneling terms.

In the limit where the first sum dominates, $t_{\b m} \gg t_{\b m'}$ $\forall \b m\in M, \b m'\in M'$,
the degeneracy is lifted due to $\b m$ quasiparticles tunneling around the $M'$ domains,
which corresponds to the case where the $M'$ regions are small and the edge is in the $M$ phase.
On the other hand, in the limit where the second sum dominates, $\tilde{t}_{\b m'} \gg t_{\b m}$, the degeneracy is lifted due to
$\b m'$ quasiparticles tunneling around the $M$ domains, which corresponds to the case where the $M$ regions are small
and the edge is in the $M'$ phase.

$H_{edge}$ above simply describes a 1D generalized quantum spin chain. One way to define this mapping
to a spin chain is as follows. Every ``site'' $i$ of the quantum spin chain
can be associated with a pair $(2i,2i+1)$  of neighboring defects. Furthermore each site $i$ of the spin chain is assigned
a $D$-dimensional Hilbert space, associated with the finite dimensional representation of the algebra
\begin{align}
W_{\b m}(a_{i}) W_{\b m'}(b_{i}) = W_{\b m'}(b_{i}) W_{\b m}(a_{i}) e^{2\pi i \b m^T K^{-1} \b m'},
\end{align}
which we studied in Sec. \ref{qdimSec} (see Fig. \ref{lineAlg} (b) for a definition of the paths $\{a_i\}$, $\{b_i\}$).
Note that the operators $W_{\b m'}(c_{2i}) = W_{\b m'}(b_i)$ (see Fig. \ref{lineAlg} (b) and Fig. \ref{transFig}) .
Also note that for the disk topology shown in Fig. \ref{transFig}, only $N-1$ pairs of defects are
independent, due to the relations:
$W_{\b m'}(c_{2N}) = \prod_{i=1}^{N-1}W_{\b m'}(c_{2i})$, and $W_{\b m}(c_{2N-1}) = \prod_{i=1}^{N-1} W_{\b m}(c_{2i-1})$.

The $A_i$ and $B_i$ in eq. (\ref{spinChainH}) satisfy:
\begin{align}
[A_i,A_j] = [B_i,B_j] = 0,
\nonumber \\
[A_i, B_j] = 0 \text{ if } j \neq i, i-1,
\nonumber \\
[A_i, B_j] \neq 0 \text{ if } j = i, i-1.
\end{align}
The latter non-vanishing commutation relation can be found by using the algebra of the Wilson line operators
$W_{\b m}(c_{2i-1})$ and $W_{\b m'}(c_{2i})$.
The $\{A_i\}$ can be viewed as the nearest neighbor spin exchange interactions
of a spin chain, while the $\{B_i\}$ act in analogy to transverse fields.

Alternatively, since the Wilson line operators are represented in the edge theory as bilinears in the parafermion
zero modes $\rgamma_{\b l;i}$, it is also possible to write the above Hamiltonian as a
``generalized parafermion chain.'' Specifically, in the edge theory
\begin{align}
W_{\b m}(c_i) &= \rgamma^\dagger_{\b m+\b m';i} \rgamma_{\b m+\b m';i+1}
\nonumber \\
&=\frac{1}{|M'|}\sum_{\b m' \in M'}  \rgamma^\dagger_{\b m+\b m';i} \rgamma_{\b m+\b m';i+1},
\end{align}
for $i$ odd. Here, $|M|$ and $|M'|$ denote the number of elements in the Lagrangian subgroups $M$ and $M'$, respectively.
The first line above holds for any $\b m' \in M'$, justifying the sum in the second line. Similarly,
\begin{align}
W_{\b m'}(c_i) &= \rgamma^\dagger_{\b m+\b m';i} \rgamma_{\b m+\b m';i+1}
\nonumber \\
&=\frac{1}{|M|} \sum_{\b m \in M} \rgamma^\dagger_{\b m+\b m';i} \rgamma_{\b m+\b m';i+1},
\end{align}
for $i$ even.
Therefore:
\begin{align}
A_i = \sum_{\b m\in M \atop \b m'\in M'} \bar{t}_{\b m} \rgamma^\dagger_{\b m+\b m';2i-1} \rgamma_{\b m+\b m';2i} + H.c.,
\nonumber \\
B_i  = \sum_{\b m'\in M' \atop \b m \in M} \bar{t}_{\b m'} \rgamma^\dagger_{\b m+\b m';2i} \rgamma_{\b m+\b m';2i+1} + H.c.,
\end{align}
where $\bar{t}_{\b m} = t_{\b m}/|M|$, and $\bar{t}_{\b m'} = t_{\b m'}/|M'|$. The algebra of the ``parafermion'' operators
$\rgamma_{\b li}$ was given in (\ref{pfAlg1}), (\ref{pfAlg2}).

As we noted previously, the commutation relations of zero modes on
defects and ``anti-defects'' (ie the defects labelled $2i$ and $2i+1$) are equivalent. Therefore, the above Hamiltonian
has an enhanced symmetry, associated with translating by one defect site, $\rgamma_{\b l, i} \rightarrow \rgamma_{\b l, i+1}$,
and replacing $\bar{t}_{\b m} \leftrightarrow \bar{t}_{\b m'}$. Such self-duality can lead to powerful constraints on the phase diagram
of the parafermion chains. In some simple cases,\cite{barkeshli2013genon,lindner2012,clarke2013} the properties of such
parafermion chains\cite{fendley2012} are well-understood, and the critical points
can, for example, give rise to parafermion CFTs.\cite{zamolodchikov1985} For example, for the simple case of ferromagnetism/superconductivity domain wall at the fractional quantum spin Hall edge\cite{lindner2012,clarke2013}, the edge theory (\ref{spinChainH}) describes $Z_m$ parafermions\cite{fendley2012}. For $m=2,3$, we know that there is only one phase transition which occurs at the self-dual point of the model, so that the self-dual Hamiltonian $\bar{t}_{\b m} =\bar{t}_{\b m'}$ must be at the critical point.

We note that another way of studying the critical phenomena associated with different gapped edges
is simply to consider the theory obtained by uniformly and simultaneously adding both sets
of backscattering terms associated with $M$ and $M'$-edges:
\begin{align}
\label{competeEdge}
\delta H_{edge} = \sum_{i=1}^N [ \lambda \cos (c_i \b m_i^T \phi ) + \lambda' \cos (c_i' \b m_i'^T \phi )].
\end{align}
When $\lambda \gg \lambda' \gg 1$, the edge will condense the Lagrangian subgroup $M$, and when
$\lambda' \gg \lambda \gg 1$ it will condense the Lagrangian subgroup $M'$. The critical phenomena associated
with the intermediate regime will generally be described by exotic conformal field theories.
In some simple cases, the behavior of such competing backscattering terms are known \cite{lecheminant2002}, and give
rise to parafermion CFTs, just as predicted by starting with the parafermion chains described above.

It would be interesting to develop a more general understanding of the possible critical phenomena
of both the generalized parafermion chains and the competing backscattering terms in eq. (\ref{competeEdge}).
Based on the simplest examples, it is natural to expect that the critical phenomena will generally be described
by a suitable class of generalized parafermion CFTs, possibly including those described in
Ref. \onlinecite{zamolodchikov1985,gepner1987,lu2010}.

We note that the arguments given above also imply that as long as the domain wall between
two kinds of gapped edges does not localize a topological zero mode, then this process of nucleating
and re-annihilating domain walls to go from one kind of edge to the other never closes the edge
energy gap. Therefore two gapped edges can always be adiabatically connected if the domain wall between them
does not localize a topological zero mode. This suggests that Lagrangian subgroups provide a complete
topological classification of gapped edges in the absence of any symmetries,
modulo any additional purely one-dimensional topological physics in fermionic systems.

\section{Conclusion}
\label{discSec}

In this paper, we have studied gapped boundaries between Abelian topological states, and the point defects
arising from junctions among different gapped edges, vastly generalizing the twist defects studied in previous
works.\cite{barkeshli2013genon,barkeshli2012a,lindner2012,clarke2013,cheng2012,barkeshli2010,bombin2010,
you2012,you2013,vaezi2013,brown2013} Using the folding process, we argued that
line defects can always be mapped to gapped boundaries between a generic topological state and the trivial
gapped state, while junctions between different line defects can be understood in terms of domain walls
between different classes of gapped edges separating a generic topological state and the trivial gapped state.

Using Abelian CS theory and its associated edge theory, we have proven, for topological phases of
both fermions and bosons, that every Lagrangian subgroup $M$ corresponds to a gapped edge
where $M$ is condensed. Edges corresponding to different Lagrangian subgroups are topologically
distinct, in the sense that there is no way to adiabatically tune from one to the other without closing
the energy gap on the edge. The physical meaning of $M$ is to determine how quasiparticles are
transmitted/reflected at the line defect. When the line defect separates two identical topological phases,
$M$ simply determines how quasiparticles are permuted by a symmetry of the topological order
as they cross the line defect.

For a special class of topologically ordered states, the $Z_N$ toric code models, we found that the gapped boundary
conditions correspond to different factorizations of $N$. We also provide an explicit microscopic lattice model
realization of each boundary condition.

We propose a topological classification of point defects, which are domain walls between different gapped edges and are characterized by
two Lagrangian subgroups, $M$ and $M'$, associated with the groups of quasiparticles that are
condensed on either side of the domain wall. We have shown that
the point defects necessarily localize topologically robust zero modes and give rise to topological
ground state degeneracies. We have derived a formula for the quantum dimension of such point defects
by studying the algebra of line operators that end on the edges. We also showed
that the localized zero modes are associated with a non-zero density of states at zero energy,
localized to the domain wall, for subsets of quasiparticles of the form $\b l = \b m + \b m'$, where $\b m\in M$,
$\b m\in M'$, $\b l\notin M, M'$.

We found that the defects can all be mapped to genons. This is defined by the property that the
Wilson line algebra that determines their topological ground state degeneracy
is equivalent to the Wilson loop algebra of some Abelian CS theory on a high genus surface, with the
genus proportional to the number of defects. This vastly generalizes the kinds of defects that can be
understood as genons, from certain class of twist defects\cite{barkeshli2013genon} to more general point
defects.

While the domain walls cannot be braided geometrically because they exist on a line defect,
we found that we can define a notion of projective non-Abelian
braiding statistics. The interactions between the edges and domain walls can be used
to realize topologically protected unitary transformations on the ground state subspace. The overall
phase of these transformations is not topologically protected, which is why it is referred to as \it projective \rm
non-Abelian braiding statistics. By studying the action of this ``braiding'' on the Wilson line operators that
define the ground state subspace, we found that the projective non-Abelian statistics can be understood in
terms of Dehn twists, or modular transformations, of the corresponding Abelian CS theory on the
high genus surface, generalizing earlier results.\cite{barkeshli2013genon}

For bosonic topologically ordered states, the topological zero modes occur only at domain walls between
gapped edges associated with different Lagrangian subgroups. Therefore for bosons,
in the absence of any symmetries, we expect that Lagrangian subgroups fully classify the topologically distinct
gapped edges.

For fermionic states, due to the possibility of Majorana zero modes protected by
fermion parity symmetry in one-dimension, two gapped edges can be topologically distinct
even if they correspond to the same Lagrangian subgroup. Therefore, the Lagrangian subgroups
classify gapped edges of fermionic systems, up to this additional $Z_2$ topological classification
arising from purely one-dimensional physics.

Finally, we pointed out that the critical phenomena between topologically distinct gapped edges can
be understood in terms of a generalized quantum spin chain or, equivalently, a generalized parafermion
chain. The properties of such chains are understood only in the simplest cases.

The study here raises many interesting questions for future research. These include understanding how to extend these
results to non-Abelian topological states, developing a deeper understanding of the critical phenomena between
different edge states, studying the interplay of this purely topological physics with global symmetries,\cite{motruk2013} understanding
the collective phenomena of the non-Abelian point defects,\cite{burrello2013} and extending results to higher dimensional topological states.

Another interesting direction is to develop a theory where the extrinsic defects
are dynamical degrees of freedom rather than static defects. In the simpler cases of twist defects,\cite{barkeshli2013genon} it was shown that
the defects can become deconfined non-Abelian excitations of a non-Abelian topological phase if the symmetry associated
with the twist defect is gauged, and the topological properties of the resulting non-Abelian state were studied in some
simple cases.\cite{barkeshli2010,barkeshli2011orb,barkeshli2010twist,barkeshli2013genon} A systematic generalization
of these results may also clarify the relation between extrinsic defects and intrinsic quasiparticle excitations.

\it Acknowledgements \rm We would like to thank Alexei Kitaev and Jason Alicea for helpful discussions, and especially
Michael Levin for introducing us to the Smith normal form of integer matrices.
This work was supported by the Simons foundation (MB), the BOCO fellowship (CMJ) and the Packard foundation (XLQ).
As this work was being completed, we became aware that the result that every Lagrangian subgroup
$M$ can correspond to a gapped edge where $M$ is condensed was independently found by M.
Levin, and included in an updated version of Ref. \onlinecite{levin2013}; our work uses results
from an early version of Ref. \onlinecite{levin2013}.

\appendix

\section{Proof of Lemma}

Here we prove the Lemma presented in Sec. \ref{proofSec}.
To prove this, we will focus on two cases independently. In the first case, $K$ has only even entries along the diagonals,
which describes topological phases where the microscopic degrees of freedom only consist of bosons. In the second case,
$K$ can have odd entries along the diagonals, which is appropriate when the microscopic degrees of freedom have at least
one species of fermions.

We note that the proof below was also presented by us recently in a shorter treatment in Ref. \onlinecite{barkeshli2013defect};
it is included here to make the paper self-contained. Our proof closely follows and builds upon
the argument in Appendix A1 of Ref. \onlinecite{levin2013} (as our work was being completed, we became aware that a
similar improved result was included in a revised draft of Ref. \onlinecite{levin2013} in Appendix A3).

\subsection{Proof for $K$ even}

Let us first consider the case where $K$ is an even matrix, meaning that its diagonal entries are all even. Note that
$K$ is also an integer symmetric non-singular matrix with vanishing signature. Consider the lattice
\begin{align}
\Gamma = \{\b m + K\b \Lambda: \b m \in M, \b \Lambda \in \mathbb{Z}^{2N}\}.
\end{align}
$\Gamma$ is a $2N$-dimensional integer lattice, and can be written as $\Gamma = U \mathbb{Z}^{2N}$, where
$U$ is a $2N$-dimensional integer matrix. Define:
\begin{align}
P = U^T K^{-1} U.
\end{align}
$P$ is an even integer symmetric matrix with unit determinant and non-vanishing signature\cite{levin2013}. The fact that it is
an even integer matrix follows because the columns of $U$ generate the Lagrangian subgroup $M$, and these
all have bosonic mutual and self-statistics by definition. The fact that it is symmetric and has vanishing signature follows
from the fact that $K$ is symmetric and has vanishing signature. Finally, $P$ has unit determinant for the following
reason. Consider any integer vector $\b \Lambda \in \mathbb{Z}^{2N}$, and any non-integer $2N$-component vector, $\b x$.
By definition of the Lagrangian subgroup, $\b{\Lambda}^T P \b x$ must be non-integer, which implies that $P \b x$ must be non-integer.
This then implies that if $P \b x$ is integer for any $2N$-component vector $\b x$, then $\b x$ must be integer, which in turn implies that
$P^{-1}$ is integer. $P$ and $P^{-1}$ can both be integer if and only if $P$ has unit determinant.

Since $P$ is an even symmetric integer matrix with vanishing signature and unit determinant,
it follows from a mathematical theorem \cite{milnor1973}
that it is always possible to find a $SL(2N;\mathbb{Z})$ transformation $W$ such that
\begin{align}
W^T P W = \left(\begin{matrix} 0 & \mathbb{I} \\ \mathbb{I} & 0 \end{matrix} \right),
\end{align}
where $I$ is an $N \times N$ identity matrix.
Thus, we consider a transformed theory:
\begin{align}
\tilde{U} &= W^T U W,
\nonumber \\
\tilde{K} &= W^T K W,
\nonumber \\
\tilde{P} &= W^T P W =  \left(\begin{matrix} 0 & \mathbb{I} \\ \mathbb{I} & 0 \end{matrix} \right)
\end{align}
Since $W \in SL(2N;\mathbb{Z})$, $\tilde{K}$ and $K$ describe topologically equivalent theories.
Clearly, the columns of $\tilde{U}$ generate the Lagrangian subgroup $M$.
Let $\tilde{\b u}_{i}$ denote the $i$th column of $\tilde{U}$. Let us extend $\tilde{K}$
to a $4N \times 4N$ matrix $K'$, which is composed of $\tilde{K}$ and $N$ copies of
$\tau_x = \left(\begin{matrix} 0 & 1 \\ 1 & 0 \end{matrix} \right)$
along the block diagonal entries:
\begin{align}
K' = \left(\begin{matrix} \tilde{K} &  & & \\
 & \tau_x &  & \\
 &  & \tau_x & \\
&  & & \ddots\end{matrix} \right),
\end{align}
where the rest of the entries are zero. Again, $K'$ describes the same
topological order as $K$. Now we define
\begin{align}
\b{m}_1'^T &= (\tilde{\b u}_1^T, 0, 1, 0, 0, \cdots,0,0)
\nonumber \\
\b{m}_2'^T &= (\tilde{\b u}_{N+1}^T, -1, 0, 0,0, \cdots,0,0)
\nonumber \\
\b{m}_3'^T &= (\tilde{\b u}_2^T,0,0,0,1,\cdots,0,0)
\nonumber \\
\b{m}_4'^T &= (\tilde{\b u}_{N+2}^T,0,0,-1,0,\cdots,0,0)
\nonumber \\
\vdots
\nonumber \\
\b{m}_{2N-1}'^T &= (\tilde{\b u}_{N}^T,0,0,\cdots,0,1)
\nonumber \\
\b{m}_{2N}'^T &= (\tilde{\b u}_{2N}^T,0,0,\cdots,-1,0)
\end{align}
Since the additional components added in $K'$ are all trivial degrees of freedom, the $2N$ vectors
$\{\b{m}_i'\}$ still generate the same Lagrangian subgroup $M$. It is easy to see that
\begin{align}
\b m_i'^T K'^{-1} \b m'_j = 0.
\end{align}
This proves the lemma for $K$ even. In practice, in most cases of interest it is easy to find $N$ columns
of $\tilde{U}$ that generate $M$ and that satisfy $\tilde{\b u}^T_i K^{-1} \tilde{\b u}_j = 0$, so the
above extension to a $4N$ dimensional $K$-matrix will not be necessary.

\subsection{Proof for $K$ odd}

Let us now consider the case where $K$ is odd (\it ie \rm it has at least one odd element along
the diagonal). As before, we define the matrix $U$, and $P = U^T K^{-1} U$. Now, $P$ is an
integer symmetric, non-singular matrix with unit determinant, non-vanishing signature, and at
least one odd element along the diagonal. Under these conditions, it is always possible to find
$W \in SL(2N;\mathbb{Z})$ such that \cite{milnor1973}
\begin{align}
W^T P W = \left(\begin{matrix} \mathbb{I} & 0 \\ 0 & -\mathbb{I} \end{matrix} \right),
\end{align}
where $I$ is an $N \times N$ identity matrix.
Thus, we consider a transformed theory:
\begin{align}
\tilde{U} &= W^T U W,
\nonumber \\
\tilde{K} &= W^T K W,
\nonumber \\
\tilde{P} &= W^T P W =  \left(\begin{matrix} \mathbb{I} & 0 \\ 0 & -\mathbb{I} \end{matrix} \right).
\end{align}
Again, the original Lagrangian subgroup is generated by the columns of $\tilde{U}$.

Let $\tilde{\b u}_{i}$ denote the $i$th column of $\tilde{U}$, and
$\tilde{\b u}_{i\pm} = \tilde{\b u}_i \pm \tilde{u}_{N+i}$, for $i = 1, \cdots, N$.

Now, as in the case where $K$ is even, let us extend the $K$-matrix to a
$4N \times 4N$ matrix $K'$, which is composed of $K$ and now with $N$ copies of
$\tau_z = \left(\begin{matrix} 1 & 0 \\ 0 & -1 \end{matrix} \right)$
along the block diagonal entries:
\begin{align}
K' = \left(\begin{matrix} \tilde{K} &  & & \\
 & \tau_z &  & \\
 &  & \tau_z & \\
&  & & \ddots\end{matrix} \right),
\end{align}
where the rest of the entries are zero. In the absence of any symmetries,
$K'$ describes the same topological order as $K$.
Now we define
\begin{align}
\b{m}_1'^T &= (\tilde{\b u}_{1+}^T, 1, 1, 0, 0, \cdots,0,0)
\nonumber \\
\b{m}_2'^T &= (\tilde{\b u}_{1-}^T, -1, 1, 0,0, \cdots,0,0)
\nonumber \\
\b{m}_3'^T &= (\tilde{\b u}_{2+}^T,0,0,1,1,\cdots,0,0)
\nonumber \\
\b{m}_4'^T &= (\tilde{\b u}_{2-}^T,0,0,-1,1,\cdots,0,0)
\nonumber \\
\vdots
\nonumber \\
\b{m}_{2N-1}'^T &= (\tilde{\b u}_{N+}^T,0,0,\cdots,1,1)
\nonumber \\
\b{m}_{2N}'^T &= (\tilde{\b u}_{N-}^T,0,0,\cdots,-1,1)
\end{align}
Since the additional components added to $K'$ are all trivial degrees of freedom, the $2N$ vectors
$\{\b{m}_i'\}$ still generate the same Lagrangian subgroup $M$. It is easy to see that
\begin{align}
\b m_i'^T K'^{-1} \b m'_j = 0.
\end{align}
This proves the lemma for $K$ odd.

\section{Ground State Degeneracy of Defect-Coupling Hamiltonians $H_{ab}$}
\label{gsdAppendix}

Here we will explain the claim made in Sec. \ref{braidSec}, that the Hamitonians $H_{12}$, $H_{1'2'}$, and $H_{22'}$, defined
in eq. (\ref{HabDef}), have $D$ degenerate ground states in the Hilbert space defined by defects $1$,$2$,$1'$,
$2'$ (see Fig. \ref{braid1}).

First, we observe that the Hilbert space defined by the defects $1$,$2$,$1'$, $2'$ is $D^2$-dimensional, and
forms a representation of the following Wilson line algebra:
\begin{align}
\label{alg}
W_{\b m}(a_i) W_{\b m'}(b_i) = W_{\b m'}(b_i) W_{\b m}(a_i) e^{\delta_{ij} i 2\pi \b m^T K^{-1} \b m'},
\end{align}
for $i,j = 1,~2,~1',~2'$, and $\b m \in M$, $\b m' \in M'$. This forms two independent copies of the same algebra,
studied in Sec. \ref{qdimSec}, each with a $D$-dimensional irreducible representation. The $D^2$ states can
each be labelled by the eigenvalues of $W_{\b m} (b_i)$:
\begin{align}
\label{basis}
W_{\b m}(b_i) |\v \alpha_1 \v \alpha_2 \rangle = e^{2\pi i \v q \cdot \v \alpha_i } |\v \alpha_1 \v \alpha_2 \rangle ,
\end{align}
where $\b m = \sum_{i=1}^N q_i \b m_i$ and $\v \alpha_i$ is a rational-valued $N$-component vector.

Let us begin by considering
\begin{align}
H_{12} &= \sum_{i=1}^N t_{i} \rgamma^\dagger_{\b l_i; 1} \rgamma_{\b l_i;2} + H.c.
\nonumber \\
&= \sum_{i=1}^N t_i W_{\b m_i}(b_1) + H.c.
\end{align}
$H_{12}$ is therefore diagonal in the basis of (\ref{basis}) and given by:
\begin{align}
H_{12} |\v \alpha_1 \v \alpha_2 \rangle = \sum_{i=1}^N 2 |t_i| \cos (2\pi (\v \alpha_1)_i + \phi_i) |\v \alpha_1 \v \alpha_2 \rangle,
\end{align}
where $t_i = |t_i| e^{i \phi_i}$. We see that $H_{12}$ is independent of the $D$ distinct eigenvalues
$\v \alpha_2$. Furthermore, unless there are accidental degeneracies, generic choices of $\phi_i$
will give distinct energies to the different eigenvalues $\v \alpha_1$, and there will be a unique
choice of $\v \alpha_1$ that minimizes the energy. Therefore, $H_{12}$ generically has $D$ distinct ground states.

The analysis for $H_{1'2'}$ is identical, except now there will be a unique choice of $\v \alpha_2$ which
minimizes the energy, while there are $D$ degenerate states associated with the possible values of $\v \alpha_1$.

Now let us consider Hamiltonians of the form $H_{11'}$ and $H_{22'}$. First consider $H_{11'}$:
\begin{align}
H_{11'} &= \sum_{i=1}^N t_i \rgamma^\dagger_{l_i;1} \rgamma_{l_i;1'} + H.c.
\nonumber \\
&=\sum_{i=1}^N t_i W_{\b m_i}(a_1) W_{\b m_i}(b_1) +H.c.
\end{align}
Let us define the operators
\begin{align}
\overline{W}_{\b m}(a_1) &\equiv W_{\b m} (a_1) W_{\b m'}(b_1),
\nonumber \\
\overline{W}_{\b m}(a_2) &\equiv W_{\b m}(a_2),
\nonumber \\
\overline{W}_{\b m'}(b_i) &\equiv W_{\b m'} (b_i) .
\end{align}
These operators satisfy the same algebra as in (\ref{alg}):
\begin{align}
\label{alg2}
\overline{W}_{\b m}(a_i) \overline{W}_{\b m'}(b_i) = \overline{W}_{\b m'}(b_i) \overline{W}_{\b m}(a_i) e^{\delta_{ij} i 2\pi \b m^T K^{-1} \b m'}.
\end{align}
In terms of these operators, $H_{11'}$ becomes
\begin{align}
H_{11'} = \sum_{i=1}^N t_i \overline{W}_{\b m_i} (a_1) + H.c.
\end{align}
Now we can directly take over the analysis of $H_{12}$ to find that $H_{11'}$ also generically has $D$ ground states.
By symmetry, the result follows for $H_{22'}$ as well.


\end{document}